\documentclass[aps,prd,twocolumn,superscriptaddress,floatfix,showpacs,showkeys]{revtex4-2}

\usepackage[utf8]{inputenc}
\usepackage[T1]{fontenc}
\usepackage{times}
\usepackage{mathrsfs}

\usepackage{amssymb,amsfonts,amsmath,amsthm,wasysym}
\usepackage{dsfont,bbm}
\usepackage{dcolumn}
\usepackage{slashed}

\usepackage{graphicx}
\usepackage{color}

\newcommand{\MS}{{\overline{\mathrm{MS}}}}
\newcommand{\mcemd}{\multicolumn{1}{c}{---}}
\newcommand*{\VmA}{V\mathord{-}A}%
\newcommand*{\BnoL}{B\mathord{\smash{\neq}}\Lambda}

\newcommand{\cO}{\mathcal {O}}

\begin{document}

\title{Updated determination of light-cone distribution amplitudes of
octet baryons in lattice QCD}

\author{G.S.~Bali}
\author{V.M.~Braun}
\author{S.~B\"urger}
\author{M.~G\"ockeler}
\email[Contact author: ]{meinulf.goeckeler@ur.de}
\author{M.~Gruber}
\author{F.~Kaiser}
\affiliation{Institut f\"ur Theoretische Physik, Universit\"at Regensburg, 93040 Regensburg, Germany}
\author{B.A.~Kniehl}
\affiliation{II.~Institut f\"ur Theoretische Physik, Universit\"at Hamburg, 22761 Hamburg, Germany}
\author{O.L.~Veretin}
\affiliation{Institut f\"ur Theoretische Physik, Universit\"at Regensburg, 93040 Regensburg, Germany}
\affiliation{II.~Institut f\"ur Theoretische Physik, Universit\"at Hamburg, 22761 Hamburg, Germany}
\author{P.~Wein}
\affiliation{Institut f\"ur Theoretische Physik, Universit\"at Regensburg, 93040 Regensburg, Germany}

\begin{abstract} 
We present updated results for the wave function normalization
constants and the first moments of the light-cone distribution
amplitudes for the lowest-lying baryon octet. The analysis is
carried out on a large number of $n_f=2+1$ lattice gauge ensembles,
including ensembles at physical pion (and kaon) masses. These are spread
across five different lattice spacings, enabling a controlled
continuum limit. The main differences with respect to our earlier
work~\cite{RQCD:2019hps} are the use of two-loop conversion factors
to an $\MS$-like scheme and of an updated set of low-energy constants in
the parametrization of the quark mass dependence. As a byproduct,
for the first time, the anomalous dimensions for local leading-twist three-quark
operators with one derivative are obtained.
\end{abstract}

\maketitle

\section{Introduction}

Light-cone distribution amplitudes~(DAs) encode the distribution of the
longitudinal momentum amongst the partons in the leading Fock states.
They are probed in hard exclusive reactions, which involve large momentum
transfer between the initial and final state hadron.

The evaluation of DAs from QCD is a challenging task, as they are genuinely
nonperturbative objects. One possibility is the calculation of moments
of DAs in lattice QCD. Results for the nucleon and hyperon DAs have
been presented in several papers, e.g., Refs.~\cite{RQCD:2019hps,Bali:2015ykx}.
For more recent discussions of other methods and the
application to nucleon form factors see Refs.~\cite{Deng:2023csv,Han:2023xbl,
Han:2023hgy,Han:2024ucv,Chen:2024fhj,Huang:2024ugd,LatticeParton:2024vck}.
The results given in Refs.~\cite{RQCD:2019hps,Bali:2015ykx} rely on the
computation of hadronic matrix elements of local three-quark operators,
which have to be renormalized. The standard renormalization
procedure consists of two steps: The nonperturbative renormalization in a
kind of momentum-subtraction scheme on the lattice is followed by a
perturbative conversion in the continuum to an $\MS$-like scheme,
which is more appropriate for phenomenological applications. Up to now,
we had to limit ourselves to one-loop accuracy in the second step,
due to the complexity of higher-loop calculations. However, it has been
observed that in several cases the perturbative expansion of the
(matrices of) conversion factors converges rather slowly.
Therefore, we reanalyze here the data presented in
Ref.~\cite{RQCD:2019hps} using two-loop conversion factors. For the
operators without derivatives these have been computed in
Ref.~\cite{Kniehl:2022ido}. The extension of these computations to
operators with one derivative is part of the present paper.
As a byproduct we obtain the anomalous dimensions for twist-three
three-quark operators with one derivative. In addition,
we take the opportunity to update the values of the low-energy constants
$F_0$, $m_b$, $D$ and $F$ used in the analysis. For our final results, see
Tables~\ref{tab.results2} and \ref{tab.moments2}. 

The paper is organized as follows. 
In Sec.~\ref{sec.DAs} we collect the relevant definitions, following
Ref.~\cite{RQCD:2019hps}. In Sec.~\ref{sec.analysis} we sketch the
analysis of our data, in particular, we discuss the changes in the
renormalization procedure due to our two-loop calculation of the conversion
matrices. We present our results in Sec.~\ref{sec.results} and compare
them with the values obtained in Ref.~\cite{RQCD:2019hps}.

Further technical details are given in the Appendices.
In Appendix~\ref{sec.3qops} we compile the three-quark operators used
in the lattice computations. The two-loop conversion matrices for
these operators are collected in Appendix~\ref{sec.converlat}.
Appendix~\ref{sec.anodimgen} is devoted to the relevant (continuum)
anomalous dimensions, in particular, we present the new two-loop contribution
to the anomalous dimensions of twist-three operators with one derivative.
The explicit form of the two-loop anomalous dimensions for our lattice
operators with one derivative can be found in Appendix~\ref{sec.anodimlat}.
A direct comparison of renormalization factors evaluated
with the help of one-loop and two-loop conversion factors is shown in
Appendix~\ref{sec.compare}.

\section{Three-quark distribution amplitudes}
\label{sec.DAs}

Baryon DAs are defined as matrix elements of renormalized three-quark
operators (we use the scheme proposed in Ref.~\cite{Krankl:2011gch})
at light-like separations:
\begin{equation}\label{eq.BDA}
\begin{split}  
&\langle 0 | \big[f_\alpha(a_1 n) g_\beta(a_2 n) h_\gamma(a_3 n)\big]
^{\mathrm R} |
B_{p,\lambda} \rangle   \\ & {} =
\frac{1}{4} \int \! [dx] \; e^{- i p \cdot n \sum_i a_i x_i}
\Big( v^B_{\alpha\beta;\gamma} V^B(x_1,x_2,x_3) \\
& \quad {} + a^B_{\alpha\beta;\gamma} A^B(x_1,x_2,x_3) +
t^B_{\alpha\beta;\gamma} T^B(x_1,x_2,x_3)+\cdots \Big) \,. 
\end{split}  
\end{equation}
On the l.h.s.\ the Wilson lines as well as the color antisymmetrization are
not written out explicitly but implied. The superscript R indicating that
the operator is renormalized will be left out again in the following.
The baryon state with momentum $p$
and helicity $\lambda$ is denoted by $| B_{p,\lambda} \rangle$, while
$\alpha,\beta,\gamma$ are Dirac indices, $n$ is a light-cone vector ($n^2=0$),
the $a_i$ are real numbers. The quark fields $f,g,h$ are of a given flavor
matching the valence-quark content of the baryon $B$.
Assuming isospin symmetry we select one representative for
each isospin multiplet, $N \equiv p$, $\Sigma \equiv \Sigma^-$, and
$\Xi \equiv \Xi^0$:
\begin{equation*}
\begin{tabular}[b]{c@{\quad}c@{\quad}c@{\quad}c}
$B$ & $f$ & $g$ & $h$ \\ \hline
$N$ & $u$ & $u$ & $d$ \\
$\Sigma$ & $d$ & $d$ & $s$ \\
$\Xi$    & $s$ & $s$ & $u$ \\
$\Lambda$ & $u$ & $d$ & $s$  
\end{tabular} \,.
\end{equation*}

On the r.h.s.\ of Eq.~\eqref{eq.BDA} the integration measure for the
longitudinal momentum fractions is given by
\begin{equation}
\int \! [dx] = \int_0^1 \!\! dx_1 \int_0^1 \!\! dx_2 \int_0^1 \!\! dx_3 \;
\delta (1-x_1-x_2-x_3)\,.
\end{equation} 
Out of the 24 terms of the general Lorentz decomposition~\cite{Braun:2000kw}
we display in Eq.~\eqref{eq.BDA} only the three leading-twist (twist three) DAs,
$V^B$, $A^B$, and $T^B$, which appear along with the structures
\begin{equation}
\begin{split}  
 v^B_{\alpha\beta;\gamma} &= (\tilde{\vphantom{n}\smash{\slashed{n}}}C)_{\alpha\beta}(\gamma_5 u^{B,+}_{p,\lambda})_\gamma\,, \\
 a^B_{\alpha\beta;\gamma} &= (\tilde{\vphantom{n}\smash{\slashed{n}}}\gamma_5C)_{\alpha\beta} (u^{B,+}_{p,\lambda})_\gamma\,, \\
 t^B_{\alpha\beta;\gamma} &= (i\sigma_{\perp \tilde n} C)_{\alpha\beta}(\gamma^\perp\gamma_5 u^{B,+}_{p,\lambda} )_\gamma\,.
\end{split}  
\end{equation}
Here $C$ is the charge conjugation matrix and we use the notation
\begin{equation}
\begin{split}  
\sigma_{\perp \tilde n}\otimes\gamma^\perp
&= \sigma^{\mu\rho} \tilde n_\rho g^\perp_{\mu\nu} \otimes \gamma^\nu \,, \\
g_{\mu\nu}^\perp &= g_{\mu\nu} -
\frac{\tilde n_\mu n_\nu+ \tilde n_\nu n_\mu}{\tilde n\cdot n} \,, \\
u^{B,+}_{p,\lambda} &= \frac{1}{2}\frac{\tilde{\vphantom{n}\smash{\slashed{n}}}\slashed{n}}{\tilde n\cdot n} u^B_{p,\lambda} \,, \\
\tilde n_\mu &= p_\mu - \frac{1}{2} \frac{m^2_B}{p\cdot n} n_\mu \,,
\end{split}  
\end{equation}
where $u^B_{\smash[t]{p,\lambda}}$ is the Dirac spinor with on-shell
momentum~$p$ and helicity~$\lambda$.

It proves convenient to define the following set of DAs:
\begin{equation} \label{eq.convenient_DAs}
\begin{split}  
\Phi_{\pm}^{B\neq\Lambda}(x_{123})&=\tfrac{1}{2} \Big([\VmA]^B(x_{123}) \pm [\VmA]^B(x_{321})\Big) \,, 
\\
 \Pi^{B\neq\Lambda}(x_{123})&= T^B (x_{132}) \,, 
\\
 \Phi_{+}^{\Lambda}(x_{123})&=\sqrt{\tfrac{1}{6}} \Big([\VmA]^\Lambda(x_{123}) + [\VmA]^\Lambda(x_{321}) \Big) \,,
\\
 \Phi_{-}^{\Lambda}(x_{123})&=-\sqrt{\tfrac{3}{2}} \Big([\VmA]^\Lambda(x_{123}) - [\VmA]^\Lambda(x_{321}) \Big) \,, 
\\
 \Pi^{\Lambda}(x_{123})&=\sqrt{6} \; T^\Lambda (x_{132}) \,, 
\end{split}  
\end{equation}
with the abbreviation $(x_{ijk})\equiv (x_i,x_j,x_k)$.
Using the phase conventions for the baryon states and the corresponding 
flavor wave functions detailed in Appendix~A of Ref.~\cite{Bali:2015ykx},
the following relations hold in the limit of SU(3) flavor symmetry
(subsequently indicated by a $\star$), where $m_u=m_d=m_s$:
\begin{equation} \label{eq.SU3_leadingtwist}
\begin{split}  
\Phi_{+}^\star &\equiv \Phi_{+}^{N\star} = \Phi_{+}^{\Sigma\star} = \Phi_{+}^{\Xi\star} = \Phi_{+}^{\Lambda\star} = \Pi^{N\star} = \Pi^{\Sigma\star} = \Pi^{\Xi\star} \,, \\
\Phi_{-}^\star &\equiv \Phi_{-}^{N\star} = \Phi_{-}^{\Sigma\star} = \Phi_{-}^{\Xi\star} = \Phi_{-}^{\Lambda\star} = \Pi^{\Lambda\star}\,.
\end{split}  
\end{equation}
In the case of SU(2) isospin symmetry,
which is exact in our $N_f=2+1$ simulation ($m_u=m_d\equiv m_\ell$) and is
only broken very mildly in the real world, the nucleon DA $\Pi^N$ is equal
to $\Phi_+^N$ in the whole $m_\ell$-$m_s$-plane.

DAs can be expanded in terms of orthogonal polynomials
$\mathcal{P}_{nk}$~\cite{Braun:2008ia}
in such a way that the coefficients have autonomous scale dependence at one
loop. For the DAs defined in Eq.~\eqref{eq.convenient_DAs}, this expansion
reads
\begin{equation} \label{eq.moments}
\begin{split}  
\Phi_{+}^{B} &= 120 \, x_1 x_2 x_3 \big( \varphi^B_{00} \mathcal P_{00} + \varphi^B_{11} \mathcal P_{11} + \cdots \big) \, , \\
 \Phi_{-}^{B} &= 120 \, x_1 x_2 x_3 \big( \varphi^B_{10} \mathcal P_{10} + \cdots \big) \,, \\
 \Pi^{B\neq\Lambda} &= 120 \, x_1 x_2 x_3 \big( \pi^B_{00} \mathcal P_{00} + \pi^B_{11} \mathcal P_{11} + \cdots \big) \, ,  \\
 \Pi^{\Lambda} &= 120 \, x_1 x_2 x_3 \big( \pi^{\Lambda}_{10} \mathcal P_{10} + \cdots \big) \,.
\end{split}  
\end{equation}
In this way all nonperturbative information is encoded in the set of
scale-dependent coefficients $\varphi^B_{nk}$, $\pi^B_{nk}$ (also called
shape parameters), which can be related to matrix elements of local operators
that are calculable on the lattice. The first few polynomials are
\begin{equation} 
\begin{split}  
\mathcal{P}_{00} &= 1 \,, \\
\mathcal{P}_{10} &= 21(x_1-x_3)\,, \\
\mathcal{P}_{11} &= 7(x_1-2x_2 + x_3)\,.
\end{split}  
\end{equation}

The leading contributions in Eq.~\eqref{eq.moments} are
$120 \, x_1 x_2 x_3 \varphi^B_{00}$ and
$120 \, x_1 x_2 x_3 \pi^{B\neq\Lambda}_{00}$. They are usually referred to as
the asymptotic DAs. The corresponding normalization coefficients
$\varphi^B_{00}$ and $\pi^{B\neq\Lambda}_{00}$
can be thought of as the wave functions at the origin. They
are also denoted as $f^B$ and $f_T^{\BnoL}$:
\begin{equation} 
f^B = \varphi^B_{00}\,, \quad f^{B\neq\Lambda}_T = \pi^B_{00}\,.
\end{equation}
Using chiral quark fields
$q^{\uparrow \downarrow} =\frac{1}{2} (\mathds 1 \pm \gamma_5) q$
and baryon spinors
$\smash{u^{B \uparrow \downarrow}_{p,\lambda}} =
\frac{1}{2}(\mathds 1 \pm \gamma_5) u^{B}_{p,\lambda}$,
they can be expressed as matrix elements of local currents with all quark
fields taken at the origin:
\begin{equation} \label{eq.norm_twist3}
\begin{split}  
\langle 0 |\big( f^{\uparrow T}C\slashed{n}g^\downarrow \big)
\slashed{n} h^\uparrow|(\BnoL)_{p,\lambda}\rangle
&= -\tfrac{1}{2}f^Bp\cdot n \slashed{n}u^{B\uparrow}_{p,\lambda} \,, \\
\langle 0 |\big( u^{\uparrow T}C\slashed{n}d^\downarrow \big)
\slashed{n} s^\uparrow|\Lambda_{p,\lambda}\rangle
&= - \sqrt{\tfrac{3}{8}} 
f^\Lambda p\cdot n \slashed{n}u^{\Lambda\uparrow}_{p,\lambda} \,, \\
\langle 0 |\big( f^{\uparrow T}C\gamma^\mu\slashed{n}g^\uparrow \big)
\gamma_\mu\slashed{n} h^\downarrow|(\BnoL)_{p,\lambda}\rangle
&= 2f_T^Bp\cdot n \slashed{n}u^{B\uparrow}_{p,\lambda} \,. 
\end{split}  
\end{equation}
In the limit of isospin symmetry these two couplings
coincide for the nucleon, $f^N_T = f^N$, see, e.g.,
Ref.~\cite{Chernyak:1983ej}.
For the $\Lambda$ baryon the zeroth moment of $T^\Lambda$
vanishes by construction so that only one leading-twist normalization
constant $f^\Lambda$ exists.

We also consider normalized first moments of $[\VmA]^B$ and~$T^{\BnoL}$,
\begin{equation} \label{eq.firstmoments}
\begin{split}  
\langle x_i \rangle^B  &= \frac{1}{f^B}\int \![dx]\, x_i\, [\VmA]^B\,,\\
\langle x_i \rangle^{\BnoL}_T &= \frac{1}{f_T^B} \int \![dx]\, x_i \, T^B\,,
\end{split}  
\end{equation}
which can be computed from the shape parameters:
\begin{equation} 
\begin{split}  
\langle x_1 \rangle^{B\neq\Lambda} &= \frac13 +
\frac13\widehat\varphi_{11}^{\,B}  + \widehat\varphi_{10}^{\,B} \,, \\
\langle x_2 \rangle^{B\neq\Lambda} &= \frac13 -
\frac23\widehat\varphi_{11}^{\,B} \,, \\
\langle x_3 \rangle^{B\neq\Lambda} &= \frac13 +
\frac13\widehat\varphi_{11}^{\,B} - \widehat\varphi_{10}^{\,B} \,,
\\
\langle x_1 \rangle^{B\neq\Lambda}_T &= \frac13 +
\frac13\widehat\pi_{11}^{\,B} \,, \\ 
\langle x_2 \rangle^{B\neq\Lambda}_T &= \frac13 +
\frac13\widehat\pi_{11}^{\,B} \,, \\
\langle x_3 \rangle^{B\neq\Lambda}_T &= \frac13 -
\frac23\widehat\pi_{11}^{\,B} \,,
\\
\langle x_1 \rangle^\Lambda &= \frac13 +
\frac13\widehat\varphi_{11}^{\,\Lambda} -
\frac13\widehat\varphi_{10}^{\,\Lambda} \,, \\ 
\langle x_2 \rangle^\Lambda &= \frac13 -
\frac23\widehat\varphi_{11}^{\,\Lambda} \,, \\
\langle x_3 \rangle^\Lambda &= \frac13 +
\frac13\widehat\varphi_{11}^{\,\Lambda} +
\frac13\widehat\varphi_{10}^{\,\Lambda} \,,
\end{split}  
\end{equation}
where 
\begin{equation} 
\widehat \varphi_{nk}^{\,B} = \frac{\varphi_{nk}^B}{f^B}
\quad , \quad  
\widehat \pi_{11}^{\,B\neq\Lambda} = \frac{\pi_{11}^B}{f^B_T} \,.
\end{equation}
Loosely speaking, one can interpret these moments as fractions of the
baryon's total momentum carried by the individual valence quarks,
e.g., $\langle x_1 \rangle^N$ corresponds to the momentum fraction
carried by the spin-up $u$ quark $u^\uparrow$ in the nucleon,
$\langle x_2 \rangle^N$ corresponds to the momentum fraction carried by
the spin-down $u$ quark $u^\downarrow$, etc. These assignments will be
indicated in the tables showing our results for these moments.

The 21 DAs of higher twist (indicated in Eq.~\eqref{eq.BDA} by the
ellipsis on the r.h.s.) only involve two new normalization constants
($\lambda_1^B$ and $\lambda_2^B$) for the isospin-nonsinglet baryons
($N$, $\Sigma$, $\Xi$) and three ($\lambda_1^\Lambda$, $\lambda_T^\Lambda$,
and $\lambda_2^\Lambda$) for the $\Lambda$~baryon. These can be defined
as matrix elements of local three-quark twist-four operators without
derivatives:
\begin{equation} \label{eq.norm_twist4}
\begin{split}  
\langle 0 | \big(f^{\uparrow T} C \gamma^\mu g^\downarrow \big) \gamma_\mu
h^\uparrow | (\BnoL) _{p,\lambda} \rangle
&= - \tfrac{1}{2} \lambda_1^B m_B u^{B\downarrow}_{p,\lambda} \,, 
\\
\langle 0 | \big(f^{\uparrow T} C \sigma^{\mu\nu} g^\uparrow \big)
\sigma_{\mu\nu} h^\uparrow | (\BnoL) _{p,\lambda} \rangle
&= \lambda_2^B m_B u^{B\uparrow}_{p,\lambda} \,, 
\\
\langle 0 | \big(u^{\uparrow T} C \gamma^\mu d^\downarrow \big) \gamma_\mu
s^\uparrow | \Lambda _{p,\lambda} \rangle
&=\tfrac{1}{2\sqrt{6}} \lambda_1^\Lambda m_\Lambda
u^{\Lambda\downarrow}_{p,\lambda} \,, 
\\
\langle 0 | \big(u^{\uparrow T} C d^\uparrow \big) s^\downarrow |
\Lambda _{p,\lambda} \rangle
&= \tfrac{1}{2\sqrt{6}} \lambda_T^\Lambda m_\Lambda
u^{\Lambda\downarrow}_{p,\lambda} \,, 
\\
\langle 0 | \big(u^{\uparrow T} C d^\uparrow \big) s^\uparrow |
\Lambda _{p,\lambda} \rangle
&= \tfrac{-1}{4\sqrt{6}} \lambda_2^\Lambda m_\Lambda
u^{\Lambda\uparrow}_{p,\lambda} \,.
\end{split}  
\end{equation}
The definitions are chosen such that in the flavor symmetric limit
\begin{equation} \label{eq.SU3_highertwist}
\begin{split}  
\lambda_1^\star &\equiv \lambda_1^{N\star} = \lambda_1^{\Sigma\star} =
\lambda_1^{\Xi\star} = \lambda_1^{\Lambda\star} = \lambda_T^{\Lambda\star} \,,
\\
\lambda_2^\star &\equiv \lambda_2^{N\star} = \lambda_2^{\Sigma\star} =
\lambda_2^{\Xi\star} = \lambda_2^{\Lambda\star} \,.
\end{split}  
\end{equation}

\section{Data analysis}
\label{sec.analysis}

We strictly follow the procedure described in Ref.~\cite{RQCD:2019hps}.
In particular, we work with the same set of gauge configurations
generated within the CLS (Coordinated Lattice Simulations)
effort~\cite{Bruno:2014jqa} with $n_f=2+1$ dynamical quarks. 
We also use the same lattice operators as in
Refs.~\cite{RQCD:2019hps,Bali:2015ykx}, collected in
Appendix~\ref{sec.3qops}, and perform the renormalization as described in
Ref.~\cite{RQCD:2019hps}. We first calculate the renormalization and mixing
coefficients nonperturbatively within a (regularization-independent)
momentum-subtraction scheme adapted to the case of three-quark operators.
These coefficients are subsequently
converted to the $\MS$-like scheme suggested in Ref.~\cite{Krankl:2011gch}
with the help of (continuum) perturbation theory. In Ref.~\cite{RQCD:2019hps}
we had only one-loop results at our disposal, while we can now make use of
our new two-loop conversion matrices (see Appendix~\ref{sec.converlat}) .
We perform several fits to the scale dependence of the
resulting numbers, extract the renormalization matrices at the target scale
of 2 GeV from each of these fits, and use them in independent analyses
of the physical quantities to obtain the final values and systematic
uncertainties of our results.
The essential elements of the fits are collected in Table~\ref{tab.choices},
which supersedes Table~1 in Ref.~\cite{RQCD:2019hps}. Here $\mu_1$ is the
initial scale of the fit range, the loop order of the conversion matrices
is given by $n_{\mathrm{loops}}$, and $n_{\mathrm{disc}}$ denotes the number
of terms in the parametrization of the lattice artifacts. As we have
updated the scales following Ref.~\cite{RQCD:2022xux}, the uncertainty
due to the scale setting is now reduced to 1.2\% (from 3\% in
Ref.~\cite{RQCD:2019hps}) and the value of $\lambda^2_{\mathrm{scale}}$
in fit 5 has been changed accordingly. Finally,
$\Lambda^{(3)}_{\MS} = 341(12) \, \mathrm{MeV}$~\cite{Bruno:2017gxd}
is varied within its uncertainty.
Also the pion masses used for the chiral extrapolations have changed slightly.
Further technical details of our renormalization procedure can be found
in Ref.~\cite{RQCD:2020kuu}.

\begin{table}[h]
\caption{\label{tab.choices} Fit choices regarding the determination of the
renormalization and mixing factors.}
\begin{ruledtabular}
\begin{tabular}{cD{.}{.}{2}ccD{.}{.}{2}c}
Fit & \multicolumn{1}{c}{$\mu_1^2 [\mathrm{GeV}^2]$} & $n_{\mathrm{loops}}$ &
  $n_{\mathrm{disc}}$ & \multicolumn{1}{c}{$\lambda^2_{\mathrm{scale}}$} &
  $\Lambda^{(3)}_{\MS} [\mathrm{MeV}]$ \\[0.1cm]
\hline \\[-0.3cm]
1 & 4  & 2 & 3 & 1.0   & 341\\
2 & 10 & 2 & 3 & 1.0   & 341\\
3 & 4  & 1 & 3 & 1.0   & 341\\
4 & 4  & 2 & 2 & 1.0   & 341\\
5 & 4  & 2 & 3 & 1.012 & 341\\
6 & 4  & 2 & 3 & 1.0   & 353\\
\end{tabular}
\end{ruledtabular}

\end{table}

\begin{figure}
\includegraphics[width=.45\textwidth]{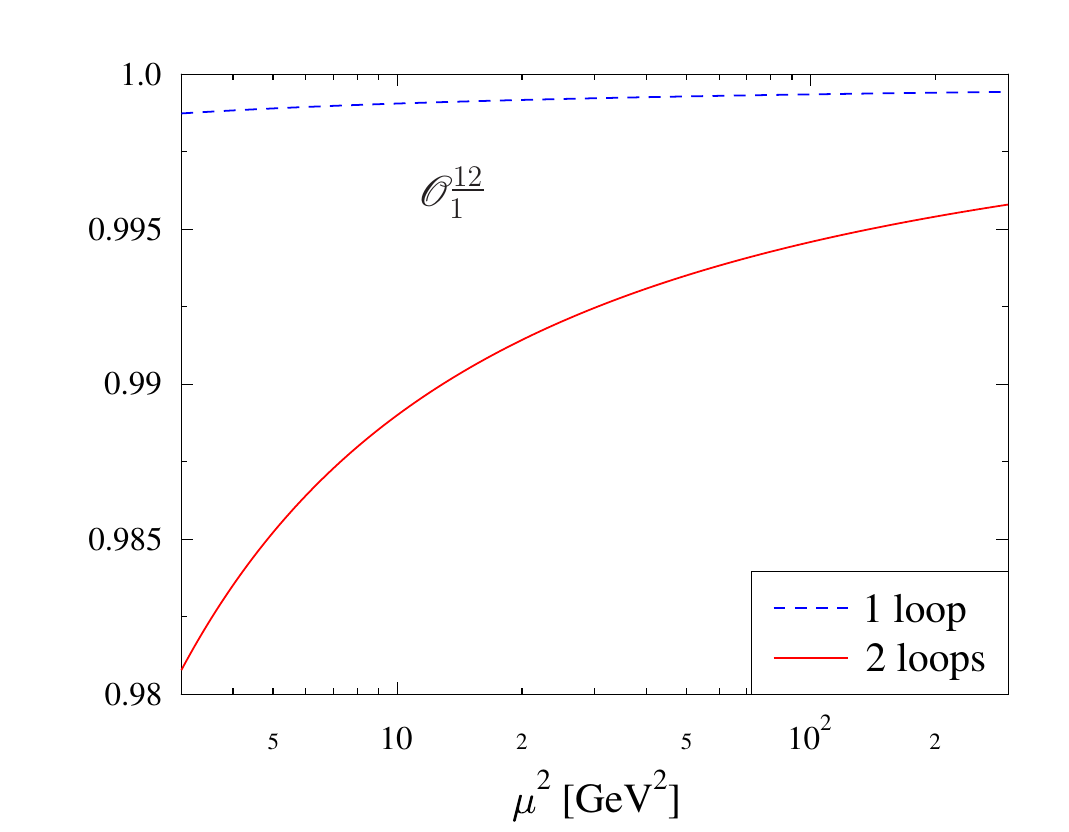}
\includegraphics[width=.45\textwidth]{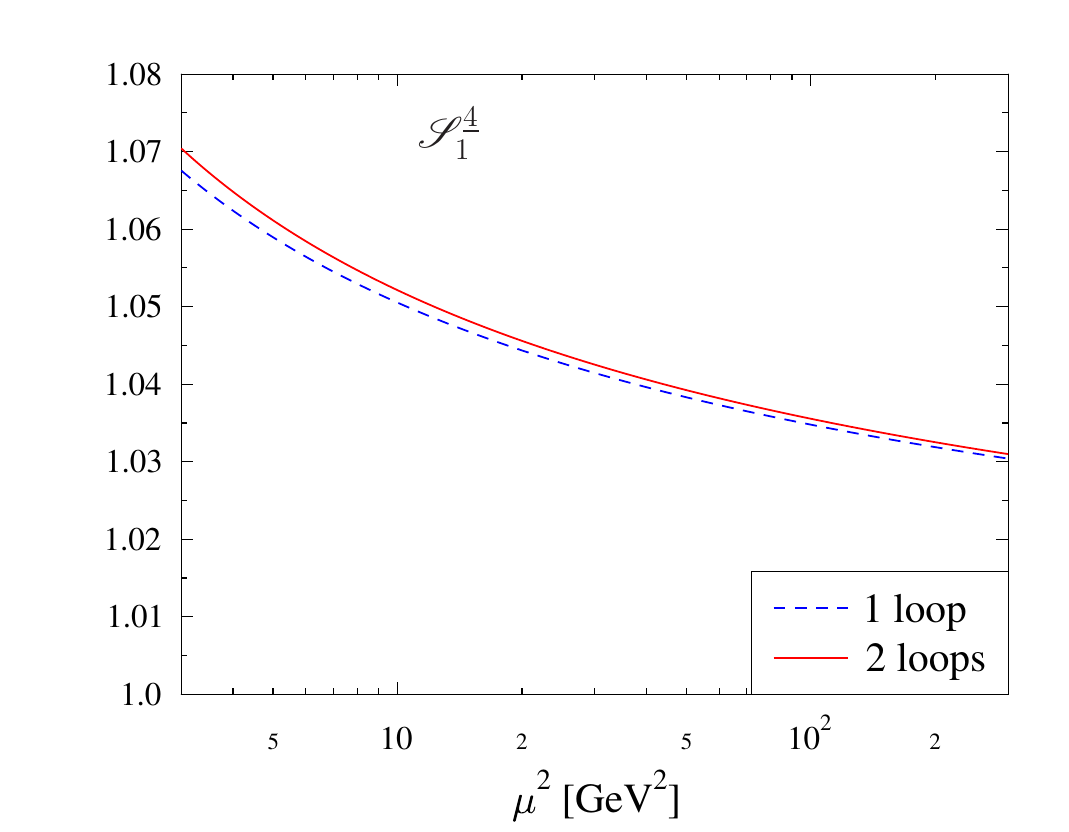}
\includegraphics[width=.45\textwidth]{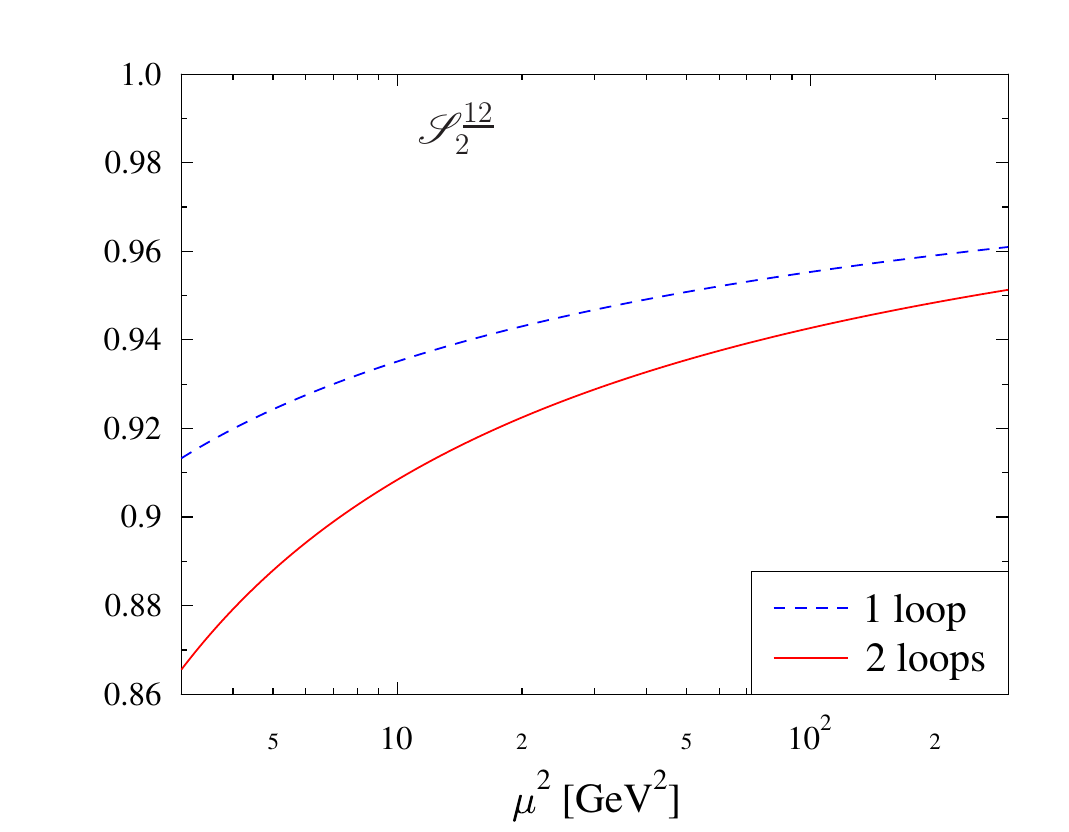}
\caption{\label{fig.conversion} Conversion factors for the multiplets
$\mathscr O_1^{\underbar{$\scriptstyle 12$}}$, 
$\mathscr S_1^{\underbar{$\scriptstyle 4$}}$, and
$\mathscr S_2^{\underbar{$\scriptstyle 12$}}$ in the one- and two-loop
approximation. The tree-level results equal $1$.}
\end{figure}

The two-loop calculation of the conversion matrices has been performed
with the help of dimensional regularization employing standard techniques.
However, for three-quark operators there are subtleties due to contributions
of evanescent operators that have to be taken into account. We employ the
renormalization scheme suggested in Ref.~\cite{Krankl:2011gch} (see
also Ref.~\cite{Gracey:2012gx}). In a first step, we use integration by parts
to reduce the Feynman integrals to a smaller set of so-called master
integrals (4 integrals at one-loop order and 44 integrals at two-loop
order). These master integrals are then evaluated
numerically. The details of the calculation in the case of the operators
without derivatives can be found in Ref.~\cite{Kniehl:2022ido}. Notice that
the number of spin tensor structures to be considered in the calculation
increases from 581 for the operators without derivatives to 2895 for the
operators with one derivative.

The uncertainties due to the numerical integration amount to
at most a few permille in the two-loop conversion coefficients.
For our final results they are completely negligible.

As a byproduct of these computations we also obtain the two-loop anomalous
dimensions of the twist-three three-quark operators with one derivative
that have not been known previously, see Appendix~\ref{sec.anodimgen}.
Notice that in the case of the operators without derivatives even the
three-loop anomalous dimensions are known~\cite{Gracey:2012gx}. For
the operators with one derivative that are used in the lattice calculations
of baryon DAs we collect the anomalous dimensions in
Appendix~\ref{sec.anodimlat}. Numerical values of the two-loop conversion
matrices for the operators and the momentum configuration employed in
Refs.~\cite{RQCD:2019hps,Bali:2015ykx,RQCD:2020kuu} can be found
in Appendix~\ref{sec.converlat}. We stress that the anomalous dimensions
beyond one-loop order and the conversion matrices starting at one loop
are scheme dependent. Our results are based on the renormalization scheme
of Ref.~\cite{Krankl:2011gch}. 

\begin{figure}
\includegraphics[width=.45\textwidth]{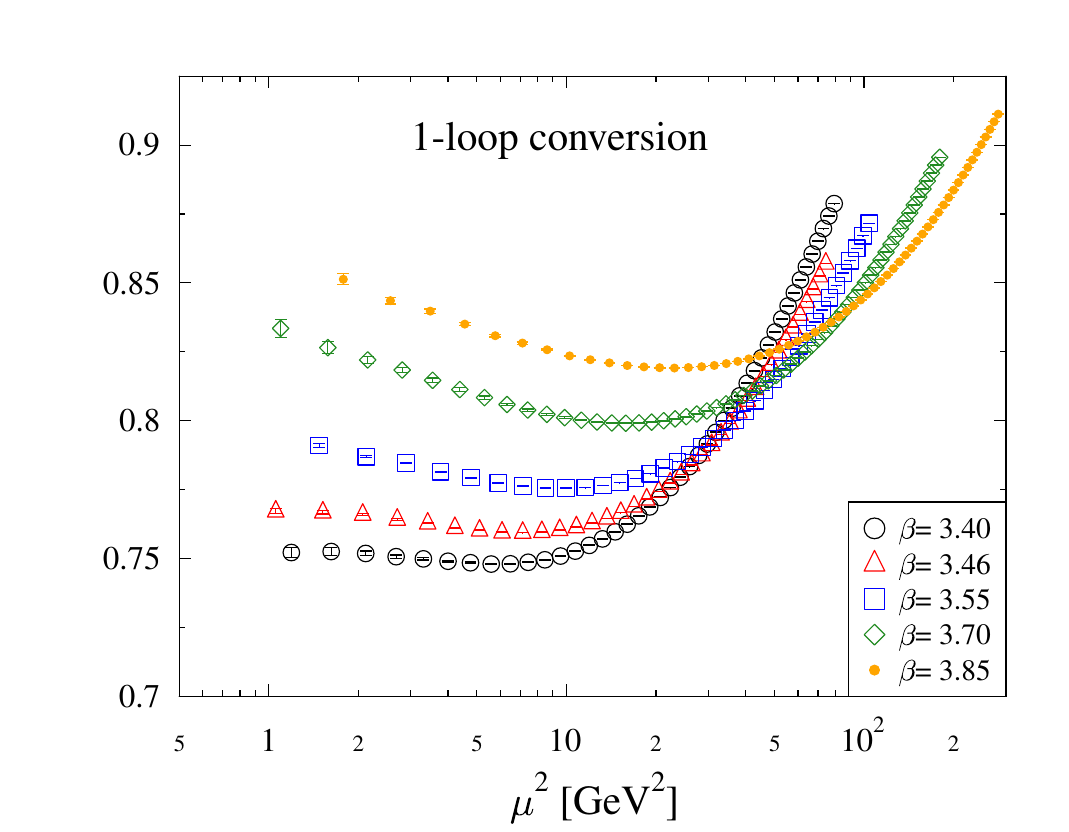}
\includegraphics[width=.45\textwidth]{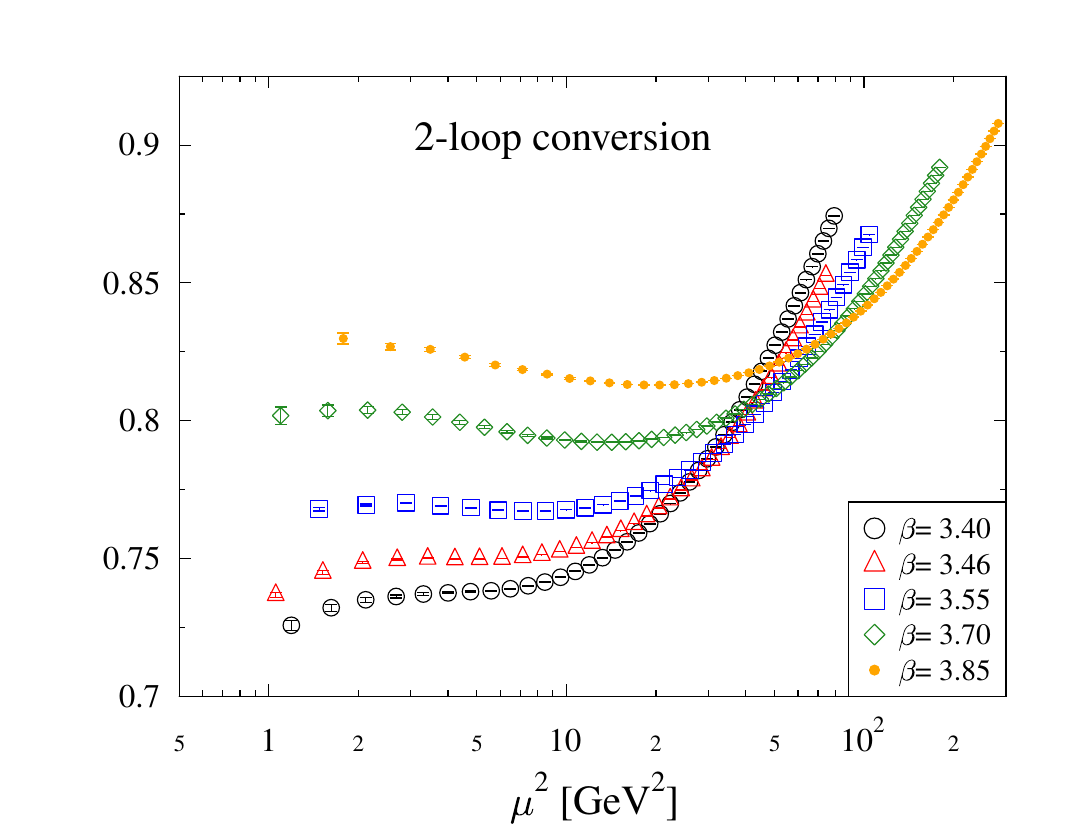}
\caption{\label{fig.zfac0d1} Renormalization factor for the multiplet
$\mathscr O_1^{\underbar{$\scriptstyle 12$}}$ rescaled to the target
scale of 2 GeV computed with the help of the one-loop (top panel)
and the two-loop (bottom panel) conversion factor. In both cases the
three-loop anomalous dimension has been used. The coupling $\beta=3.4 (3.85)$
corresponds to the coarsest (finest) lattice spacing $a=0.085
(0.039)$~fm, employed here, see Refs.~\cite{RQCD:2019hps,RQCD:2022xux}.}
\end{figure}

\begin{figure}
\includegraphics[width=.45\textwidth]{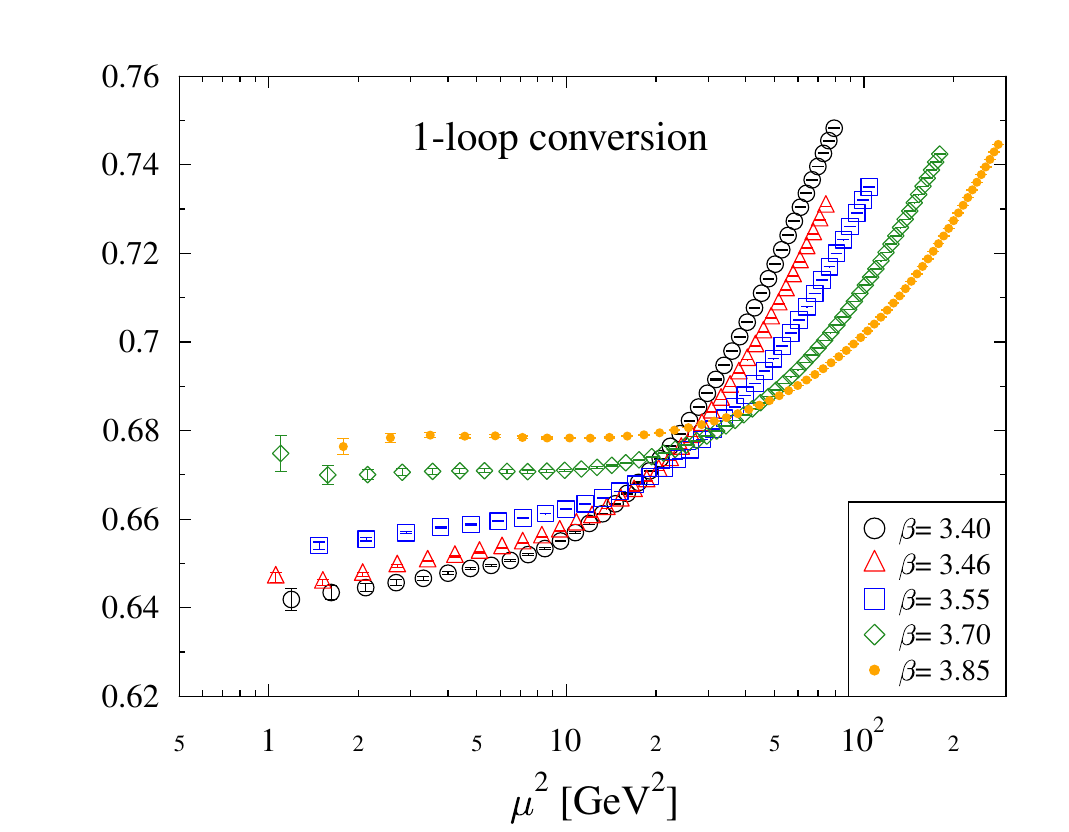}
\includegraphics[width=.45\textwidth]{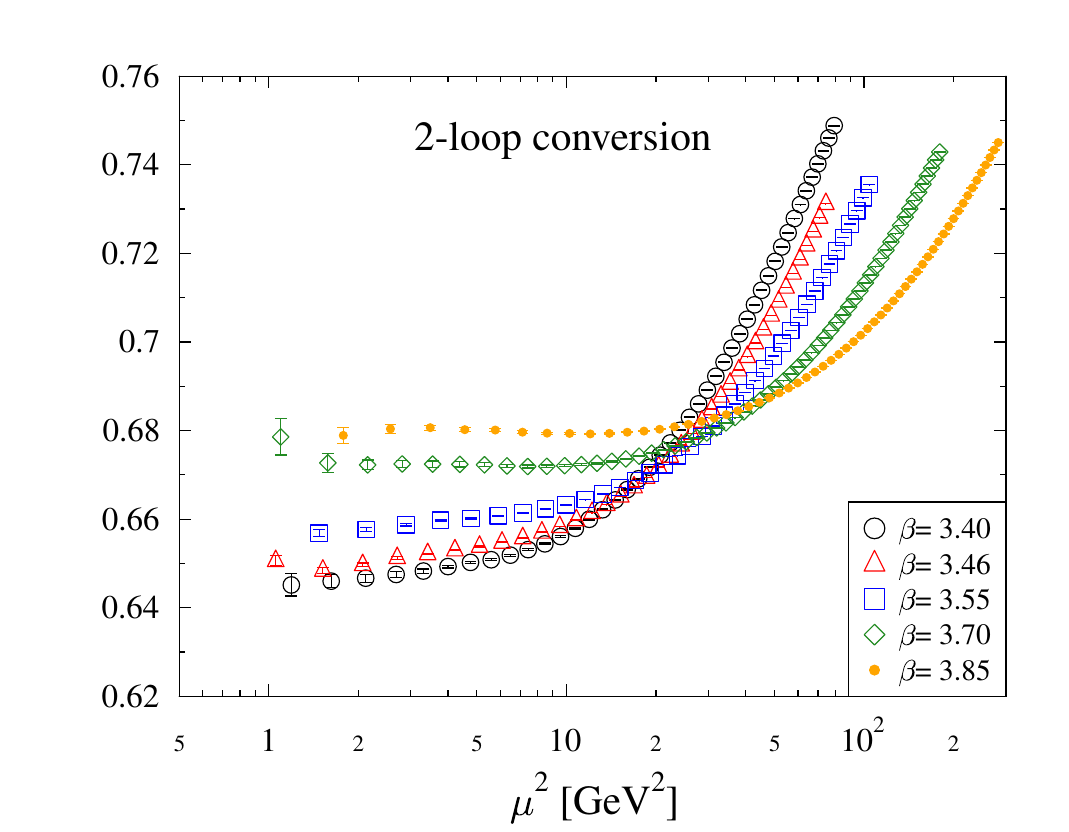}
\caption{\label{fig.zfac0d4} Renormalization factor for the multiplet
$\mathscr S_1^{\underbar{$\scriptstyle 4$}}$ rescaled to the target scale
of 2 GeV computed with the help of the one-loop (top panel) and the two-loop
(bottom panel) conversion factor. In both cases the three-loop anomalous
dimension has been used.}
\end{figure}

\begin{figure}
\includegraphics[width=.45\textwidth]{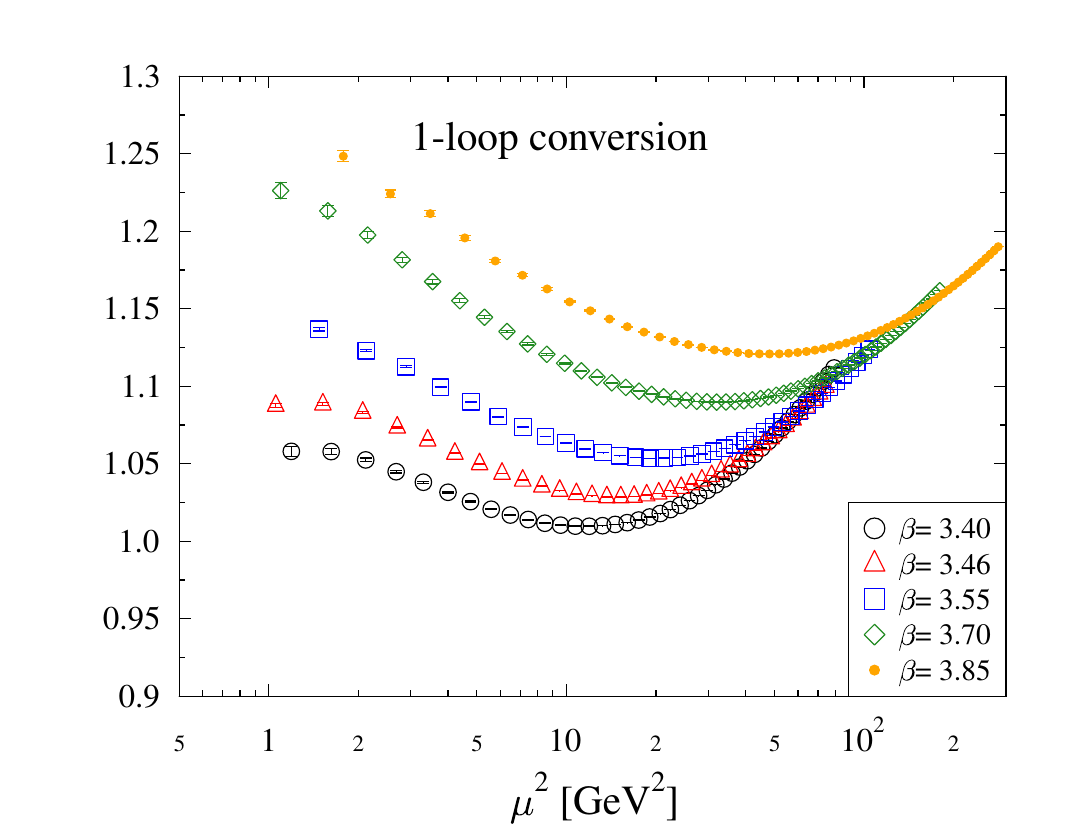}
\includegraphics[width=.45\textwidth]{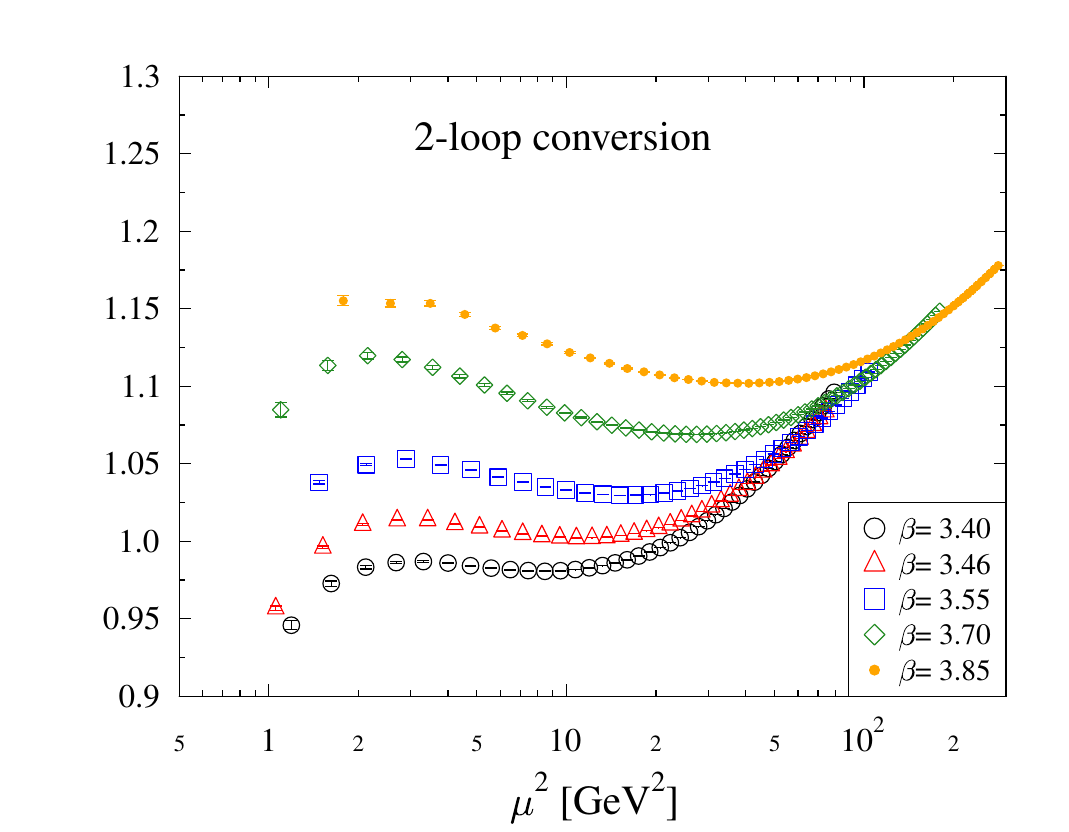}
\caption{\label{fig.zfac1d16} Renormalization factor for the multiplet
$\mathscr S_2^{\underbar{$\scriptstyle 12$}}$ rescaled to the target scale
of 2 GeV computed with the help of the one-loop (top panel) and the two-loop
(bottom panel) conversion factor. In both cases the two-loop anomalous
dimension has been used.}
\end{figure}

While in most cases the perturbative expansion of the conversion matrices
shows the expected behavior with a two-loop correction that is smaller than
the one-loop contribution for reasonably large scales, there are a few
cases where either the one-loop term is unusually small and the two-loop
term is of a reasonable size or the two-loop correction is considerably
smaller than the one-loop contribution. In Fig.~\ref{fig.conversion} we
show examples for these two situations (operators without derivatives
$\mathscr O_1^{\underbar{$\scriptstyle 12$}}$
and $\mathscr S_1^{\underbar{$\scriptstyle 4$}}$ in the notation of
Ref.~\cite{RQCD:2020kuu}, see also Appendix~\ref{sec.3qops}) as well as
for the usual behavior (multiplet
$\mathscr S_2^{\underbar{$\scriptstyle 12$}}$ of operators with one derivative).

In all the cases, the scale dependence of the renormalization and
mixing coefficients obtained with two-loop conversion looks rather
satisfactory in the sense that after rescaling to the target scale
of 2 GeV the data tend to develop approximate
plateaus as the lattice spacing decreases (i.e.\ as $\beta$ increases).
(Notice, however, the cautionary remarks at the end of Sec.~VII in
Ref.~\cite{RQCD:2020kuu}.) This can be seen in
Figs.~\ref{fig.zfac0d1} -- \ref{fig.zfac1d16}, where we display the resulting
renormalization factors evaluated either with the one-loop or the two-loop
conversion factor. Given the conversion factors shown in the middle panel of
Fig.~\ref{fig.conversion} it is not surprising that the two plots in
Fig.~\ref{fig.zfac0d4} are hardly distinguishable, where the upper plot
shows the same data as Fig.~4 in Ref.~\cite{RQCD:2019hps}.
For an easier comparison of the effect of one-loop and two-loop
conversion we plot some of the data presented in Figs.~\ref{fig.zfac0d1} and
\ref{fig.zfac1d16} in a different form in Appendix~\ref{sec.compare}.

\section{Results}
\label{sec.results}

\begin{table*}
\caption{\label{tab.results0}
Results for the couplings and shape parameters using two-loop conversion
factors, compared to the values obtained with one-loop conversion from
Ref.~\cite{RQCD:2019hps}. For this comparison we employ the ``old''
values of the low-energy constants taken from Ref.~\cite{Ledwig:2014rfa}.
All quantities are given in units of $10^{-3} \mathrm{GeV}^2$ in our $\MS$-like
scheme at a scale $\mu = 2 \, \mathrm{GeV}$ with three active quark flavors.
The superscripts and subscripts denote the statistical error after extrapolation
to the physical point. The numbers in parentheses give estimates for the
systematic errors due to renormalization~($r$), continuum extrapolation~($a$),
and chiral extrapolation~($m$).
Due to the scale setting uncertainty the new (old) results carry an additional
error of 1.2\% (3\%), which is not displayed.}
\begin{ruledtabular}
\begin{tabular}{cD{.}{.}{22}D{.}{.}{22}D{.}{.}{22}D{.}{.}{22}}
$B$ & \multicolumn{1}{c}{$N$} & \multicolumn{1}{c}{$\Sigma$} & \multicolumn{1}{c}{$\Xi$} & \multicolumn{1}{c}{$\Lambda$}\\
\hline \\[-0.2cm]
$f^B$               & 3.55^{+6}_{-4}(1)_r(0)_a(0)_m            & 5.32^{+5}_{-5}(2)_r(0)_a(5)_m            & 6.14^{+7}_{-7}(2)_r(0)_a(15)_m           & 4.89^{+7}_{-4}(2)_r(0)_a(6)_m           \\
\cite{RQCD:2019hps} & 3.54^{+6}_{-4}(1)_r(2)_a(0)_m            & 5.31^{+5}_{-4}(1)_r(3)_a(4)_m            & 6.11^{+7}_{-6}(2)_r(4)_a(13)_m           & 4.87^{+7}_{-4}(2)_r(3)_a(5)_m           \\
$f_T^B$             & 3.55^{+6}_{-4}(1)_r(0)_a(0)_m            & 5.15^{+5}_{-4}(2)_r(0)_a(5)_m            & 6.32^{+8}_{-7}(2)_r(0)_a(16)_m           & \mcemd                                  \\
\cite{RQCD:2019hps} & 3.54^{+6}_{-4}(1)_r(2)_a(0)_m            & 5.14^{+5}_{-4}(1)_r(3)_a(3)_m            & 6.29^{+8}_{-7}(1)_r(4)_a(15)_m           & \mcemd                                  \\
$\varphi_{11}^B$    & 0.125^{+6}_{-5}(7)_r(23)_a(0)_m          & 0.203^{+4}_{-6}(9)_r(44)_a(0)_m          & -0.001^{+4}_{-3}(12)_r(0)_a(0)_m         & 0.250^{+8}_{-7}(9)_r(49)_a(0)_m         \\
\cite{RQCD:2019hps} & 0.118^{+6}_{-5}(8)_r(21)_a(0)_m          & 0.195^{+4}_{-6}(10)_r(40)_a(0)_m         & -0.014^{+4}_{-3}(18)_r(3)_a(0)_m         & 0.243^{+8}_{-7}(9)_r(46)_a(0)_m         \\
$\pi_{11}^B$        & 0.125^{+6}_{-5}(7)_r(23)_a(0)_m          & -0.077^{+2}_{-2}(11)_r(16)_a(0)_m        & 0.405^{+7}_{-9}(9)_r(86)_a(0)_m          & \mcemd                                  \\
\cite{RQCD:2019hps} & 0.118^{+6}_{-5}(8)_r(21)_a(0)_m          & -0.090^{+3}_{-2}(17)_r(18)_a(0)_m        & 0.399^{+7}_{-9}(9)_r(81)_a(0)_m          & \mcemd                                  \\
$\varphi_{10}^B$    & 0.184^{+20}_{-14}(2)_r(8)_a(1)_m         & 0.091^{+12}_{-24}(2)_r(5)_a(0)_m         & 0.351^{+19}_{-21}(5)_r(20)_a(1)_m        & 0.606^{+21}_{-27}(7)_r(30)_a(2)_m       \\
\cite{RQCD:2019hps} & 0.182^{+20}_{-14}(6)_r(4)_a(1)_m         & 0.090^{+11}_{-31}(3)_r(3)_a(1)_m         & 0.350^{+18}_{-20}(11)_r(11)_a(1)_m       & 0.610^{+23}_{-28}(18)_r(16)_a(2)_m      \\
$\pi_{10}^B$        & \mcemd                                   & \mcemd                                   & \mcemd                                   & 0.218^{+32}_{-27}(3)_r(12)_a(2)_m       \\
\cite{RQCD:2019hps} & \mcemd                                   & \mcemd                                   & \mcemd                                   & 0.214^{+33}_{-26}(7)_r(6)_a(2)_m        \\[0.1cm]
\hline \\[-0.2cm]
$\lambda_1^B$       & -45.5^{+1.1}_{-1.0}(0.3)_r(2.8)_a(0.6)_m & -46.9^{+0.9}_{-0.9}(0.3)_r(2.8)_a(0.5)_m & -50.5^{+0.8}_{-1.0}(0.3)_r(3.3)_a(0.3)_m & -42.9^{+0.8}_{-0.8}(0.4)_r(2.5)_a(0.5)_m\\
\cite{RQCD:2019hps} & -44.9^{+1.2}_{-0.9}(0.9)_r(3.9)_a(0.6)_m & -46.1^{+1.0}_{-0.9}(0.9)_r(4.1)_a(0.5)_m & -49.8^{+0.8}_{-0.9}(1.0)_r(4.7)_a(0.2)_m & -42.2^{+0.8}_{-0.8}(0.9)_r(3.7)_a(0.4)_m\\
$\lambda_T^B$       & \mcemd                                   & \mcemd                                   & \mcemd                                   & -53.1^{+1.1}_{-1.1}(0.4)_r(3.4)_a(0.4)_m\\
\cite{RQCD:2019hps} & \mcemd                                   & \mcemd                                   & \mcemd                                   & -52.3^{+1.2}_{-1.0}(1.1)_r(4.7)_a(0.3)_m\\
$\lambda_2^B$       & 95.0^{+2.3}_{-2.3}(0.5)_r(3.2)_a(1.4)_m  & 86.7^{+1.9}_{-1.7}(0.2)_r(3.0)_a(1.2)_m  & 100.9^{+2.4}_{-1.8}(0.3)_r(3.7)_a(1.1)_m & 100.3^{+2.5}_{-2.2}(0.6)_r(3.5)_a(1.3)_m\\
\cite{RQCD:2019hps} & 93.4^{+2.3}_{-2.2}(1.7)_r(3.7)_a(1.2)_m  & 85.2^{+1.8}_{-1.7}(1.6)_r(3.4)_a(1.3)_m  & 99.5^{+2.3}_{-1.8}(1.9)_r(4.3)_a(1.1)_m  & 98.9^{+2.4}_{-2.1}(1.9)_r(4.1)_a(1.1)_m \\
\end{tabular}
\end{ruledtabular}

\end{table*}

\begin{table}
\caption{\label{tab.moments0}
Results (using two-loop conversion factors) for the normalized first
moments~\eqref{eq.firstmoments} of the DAs $[\VmA]^B$ and~$T^{\BnoL}$,
compared to the values obtained with one-loop conversion from
Ref.~\cite{RQCD:2019hps}. For this comparison we employ the ``old''
values of the low-energy constants taken from Ref.~\cite{Ledwig:2014rfa}.
The results refer to our $\MS$-like scheme at a scale
$\mu = 2 \, \mathrm{GeV}$ . All uncertainties from our
calculation have been added in quadrature.}
\begin{ruledtabular}
\begin{tabular}{c r@{\hspace{.5em}}l r@{\hspace{.5em}}l r@{\hspace{.5em}}l r@{\hspace{.5em}}l}
$B$ & \multicolumn{2}{c}{$N$} & \multicolumn{2}{c}{$\Sigma$} & \multicolumn{2}{c}{$\Xi$} & \multicolumn{2}{c}{$\Lambda$}\\
\hline \\[-0.2cm]
$\langle x_1 \rangle^B$     & $u^\uparrow$   & $0.397_{-6}^{+7}$ & $d^\uparrow$   & $0.363_{-6}^{+4}$ & $s^\uparrow$   & $0.391_{-5}^{+5}$ & $u^\uparrow$   & $0.309_{-2}^{+2}$\\
\cite{RQCD:2019hps}         & $u^\uparrow$   & $0.396_{-6}^{+7}$ & $d^\uparrow$   & $0.363_{-7}^{+4}$ & $s^\uparrow$   & $0.390_{-4}^{+4}$ & $u^\uparrow$   & $0.308_{-3}^{+3}$\\
$\langle x_2 \rangle^B$     & $u^\downarrow$ & $0.310_{-5}^{+5}$ & $d^\downarrow$ & $0.308_{-6}^{+6}$ & $s^\downarrow$ & $0.333_{-1}^{+1}$ & $d^\downarrow$ & $0.299_{-7}^{+7}$\\
\cite{RQCD:2019hps}         & $u^\downarrow$ & $0.311_{-5}^{+5}$ & $d^\downarrow$ & $0.309_{-5}^{+5}$ & $s^\downarrow$ & $0.335_{-2}^{+2}$ & $d^\downarrow$ & $0.300_{-7}^{+7}$\\
$\langle x_3 \rangle^B$     & $d^\uparrow$   & $0.293_{-5}^{+4}$ & $s^\uparrow$   & $0.329_{-3}^{+5}$ & $u^\uparrow$   & $0.276_{-5}^{+5}$ & $s^\uparrow$   & $0.392_{-6}^{+6}$\\
\cite{RQCD:2019hps}         & $d^\uparrow$   & $0.293_{-6}^{+5}$ & $s^\uparrow$   & $0.329_{-3}^{+6}$ & $u^\uparrow$   & $0.275_{-5}^{+5}$ & $s^\uparrow$   & $0.392_{-5}^{+5}$\\[0.1cm]
\hline \\[-0.2cm]
$\langle x_1 \rangle^B_T$   & $u^\uparrow$   & $0.345_{-2}^{+2}$ & $d^\uparrow$   & $0.328_{-1}^{+1}$ & $s^\uparrow$   & $0.355_{-5}^{+5}$ & \multicolumn{2}{c}{---}\\
\cite{RQCD:2019hps}         & $u^\uparrow$   & $0.344_{-2}^{+2}$ & $d^\uparrow$   & $0.327_{-2}^{+2}$ & $s^\uparrow$   & $0.354_{-5}^{+5}$ & \multicolumn{2}{c}{---}\\
$\langle x_2 \rangle^B_T$   & $u^\uparrow$   & $0.345_{-2}^{+2}$ & $d^\uparrow$   & $0.328_{-1}^{+1}$ & $s^\uparrow$   & $0.355_{-5}^{+5}$ & \multicolumn{2}{c}{---}\\
\cite{RQCD:2019hps}         & $u^\uparrow$   & $0.344_{-2}^{+2}$ & $d^\uparrow$   & $0.327_{-2}^{+2}$ & $s^\uparrow$   & $0.354_{-5}^{+5}$ & \multicolumn{2}{c}{---}\\
$\langle x_3 \rangle^B_T$   & $d^\downarrow$ & $0.310_{-5}^{+5}$ & $s^\downarrow$ & $0.343_{-2}^{+2}$ & $u^\downarrow$ & $0.291_{-9}^{+9}$ & \multicolumn{2}{c}{---}\\
\cite{RQCD:2019hps}         & $d^\downarrow$ & $0.311_{-5}^{+5}$ & $s^\downarrow$ & $0.345_{-3}^{+3}$ & $u^\downarrow$ & $0.291_{-9}^{+9}$ & \multicolumn{2}{c}{---}\\
\end{tabular}
\end{ruledtabular}

\end{table}

\begin{table*}
\caption{\label{tab.results2}
Final results for the couplings and shape parameters, using the low-energy
constants from Ref.~\cite{RQCD:2022xux} and two-loop conversion factors.
All values are given in units of $10^{-3} \mathrm{GeV}^2$ in our $\MS$-like
scheme at a scale $\mu = 2 \, \mathrm{GeV}$ with three active quark flavors.
The superscripts and subscripts denote the statistical error after extrapolation
to the physical point. The values in parentheses give estimates for the
systematic errors due to renormalization~($r$), continuum extrapolation~($a$),
and chiral extrapolation~($m$). The latter error is estimated according
to the modified procedure described in the text. Due to the scale setting
uncertainty all results carry an additional error of 1.2\% (not displayed).}
\begin{ruledtabular}
\begin{tabular}{lD{.}{.}{22}D{.}{.}{22}D{.}{.}{22}D{.}{.}{22}}
$B$ & \multicolumn{1}{c}{$N$} & \multicolumn{1}{c}{$\Sigma$} & \multicolumn{1}{c}{$\Xi$} & \multicolumn{1}{c}{$\Lambda$}\\
\hline \\[-0.2cm]
$f^B$            & 3.29^{+6}_{-4}(1)_r(1)_a(26)_m           & 5.32^{+5}_{-5}(2)_r(2)_a(9)_m            & 6.15^{+8}_{-7}(2)_r(3)_a(25)_m           & 4.75^{+6}_{-4}(2)_r(2)_a(16)_m          \\
$f_T^B$          & 3.29^{+6}_{-4}(1)_r(1)_a(26)_m           & 5.15^{+5}_{-4}(2)_r(2)_a(9)_m            & 6.34^{+8}_{-8}(3)_r(3)_a(26)_m           & \mcemd                                  \\
$\varphi_{11}^B$ & 0.118^{+6}_{-4}(7)_r(21)_a(7)_m          & 0.200^{+4}_{-6}(10)_r(42)_a(3)_m         & -0.001^{+4}_{-3}(12)_r(0)_a(0)_m         & 0.242^{+8}_{-6}(9)_r(46)_a(8)_m         \\
$\pi_{11}^B$     & 0.118^{+6}_{-4}(7)_r(21)_a(7)_m          & -0.080^{+2}_{-2}(12)_r(16)_a(3)_m        & 0.399^{+7}_{-9}(9)_r(82)_a(6)_m          & \mcemd                                  \\
$\varphi_{10}^B$ & 0.181^{+19}_{-15}(2)_r(7)_a(3)_m         & 0.094^{+10}_{-42}(2)_r(4)_a(3)_m         & 0.361^{+16}_{-27}(5)_r(19)_a(10)_m       & 0.563^{+33}_{-27}(6)_r(27)_a(43)_m      \\
$\pi_{10}^B$     & \mcemd                                   & \mcemd                                   & \mcemd                                   & 0.237^{+30}_{-35}(4)_r(12)_a(19)_m      \\[0.1cm]
\hline \\[-0.2cm]
$\lambda_1^B$    & -44.3^{+1.0}_{-1.0}(0.3)_r(3.3)_a(1.4)_m & -47.6^{+0.9}_{-0.9}(0.3)_r(3.7)_a(1.1)_m & -50.2^{+0.7}_{-0.9}(0.3)_r(4.1)_a(0.8)_m & -44.0^{+0.7}_{-0.8}(0.5)_r(3.3)_a(1.3)_m\\
$\lambda_T^B$    & \mcemd                                   & \mcemd                                   & \mcemd                                   & -54.2^{+1.1}_{-1.1}(0.3)_r(4.3)_a(1.4)_m\\
$\lambda_2^B$    & 97.2^{+1.9}_{-2.7}(1.2)_r(1.2)_a(2.4)_m  & 85.8^{+2.0}_{-1.4}(0.8)_r(0.4)_a(2.2)_m  & 101.4^{+2.3}_{-1.9}(0.2)_r(1.1)_a(1.3)_m & 99.7^{+2.3}_{-2.4}(0.6)_r(1.2)_a(1.3)_m \\
\end{tabular}
\end{ruledtabular}

\end{table*}

\begin{table}
\caption{\label{tab.moments2}
Final results for the normalized first moments~\eqref{eq.firstmoments}
of the DAs $[\VmA]^B$ and~$T^{\BnoL}$, using the low-energy
constants from Ref.~\cite{RQCD:2022xux} and two-loop conversion factors.
The results refer to our $\MS$-like scheme at a scale
$\mu = 2 \, \mathrm{GeV}$. All uncertainties from our
calculation have been added in quadrature, including the differences with
the central values given in Table~\ref{tab.moments0}.}
\begin{ruledtabular}
\begin{tabular}{c r@{\hspace{.5em}}l r@{\hspace{.5em}}l r@{\hspace{.5em}}l r@{\hspace{.5em}}l}
$B$ & \multicolumn{2}{c}{$N$} & \multicolumn{2}{c}{$\Sigma$} & \multicolumn{2}{c}{$\Xi$} & \multicolumn{2}{c}{$\Lambda$}\\
\hline \\[-0.2cm]
$\langle x_1 \rangle^B$   & $u^\uparrow$   & $0.400_{-8}^{+9}$ & $d^\uparrow$   & $0.363_{-9}^{+4}$ & $s^\uparrow$   & $0.392_{-6}^{+5}$ & $u^\uparrow$   & $0.311_{-4}^{+3}$\\
$\langle x_2 \rangle^B$   & $u^\downarrow$ & $0.309_{-5}^{+5}$ & $d^\downarrow$ & $0.308_{-6}^{+6}$ & $s^\downarrow$ & $0.333_{-1}^{+1}$ & $d^\downarrow$ & $0.299_{-7}^{+7}$\\
$\langle x_3 \rangle^B$   & $d^\uparrow$   & $0.290_{-7}^{+6}$ & $s^\uparrow$   & $0.328_{-3}^{+8}$ & $u^\uparrow$   & $0.275_{-5}^{+6}$ & $s^\uparrow$   & $0.390_{-6}^{+6}$\\
\hline \\[-0.2cm]
$\langle x_1 \rangle^B_T$ & $u^\uparrow$   & $0.345_{-2}^{+2}$ & $d^\uparrow$   & $0.328_{-1}^{+1}$ & $s^\uparrow$   & $0.354_{-5}^{+5}$ & \multicolumn{2}{c}{---}\\
$\langle x_2 \rangle^B_T$ & $u^\uparrow$   & $0.345_{-2}^{+2}$ & $d^\uparrow$   & $0.328_{-1}^{+1}$ & $s^\uparrow$   & $0.354_{-5}^{+5}$ & \multicolumn{2}{c}{---}\\
$\langle x_3 \rangle^B_T$ & $d^\downarrow$ & $0.309_{-5}^{+5}$ & $s^\downarrow$ & $0.344_{-3}^{+3}$ & $u^\downarrow$ & $0.291_{-9}^{+9}$ & \multicolumn{2}{c}{---}\\
\end{tabular}
\end{ruledtabular}

\end{table}

Presenting our results, we begin with the numbers obtained
with the two-loop conversion matrices (factors) and the values for the
low-energy constants $F_0$, $m_b$, $D$ and $F$ used in the original
article~\cite{RQCD:2019hps}:
$F_0 = 87 \, \mathrm{MeV}$, $D=0.623$, $F=0.441$, and
$m_b = 880 \, \mathrm{MeV}$ (taken from Ref.~\cite{Ledwig:2014rfa}).
Hence the only difference with the old results is the
use of the two-loop conversion instead of the one-loop approximation.

We collect our results in Tables~\ref{tab.results0} and \ref{tab.moments0}.
The comparison with the results of Ref.~\cite{RQCD:2019hps}
in these tables highlights the effect of increasing the loop order of the
conversion. Recall that (as in Ref.~\cite{RQCD:2019hps}) the systematic
error due to the chiral extrapolation has been estimated by including
higher-order terms in the chiral expansion. Table~\ref{tab.results0}
(Table~\ref{tab.moments0}) corresponds to Table~2 (Table~4) of
Ref.~\cite{RQCD:2019hps}.

Comparing old and new results one observes that
replacing the one-loop conversion matrices with the corresponding
two-loop approximations does not lead to significant changes in our results.
Therefore the perturbative uncertainties for the quantities
considered here seem to be under control.
While the central values change only marginally, the
estimates of the errors due to the renormalization ($r$) are in general
reduced, sometimes significantly, as one could have anticipated. 
Somewhat surprising is the effect seen in the errors related to
the continuum extrapolation ($a$). For the coupling constants, i.e., for
$f^B$, $f_T^B$, $\lambda_1^B$ , $\lambda_T^B$, and $\lambda_2^B$, these
uncertainties are reduced. However, for most of the shape parameters
they are increased, in some cases even considerably. Perhaps this different
behavior is related to the fact that the coupling constants are computed
from operators without derivatives
(cf.\ Eqs.~\eqref{eq.norm_twist3} and \eqref{eq.norm_twist4}),
while the evaluation of the shape parameters involves operators with
one derivative, which are generally more difficult to handle.

At first sight, it might be surprising that the differences between the
results obtained with one-loop and two-loop conversion are smaller than
one might have expected from, e.g., Fig.~\ref{fig.zfac1d16}. However,
one has to take into account that the values of the renormalization
coefficients which we finally use are not simply read off from curves
such as those shown in Figs.~\ref{fig.zfac0d1} -- \ref{fig.zfac1d16},
but result from fits to the scale dependence. Furthermore, because SU(3) flavor
symmetry is broken, renormalization coefficients of different operators
(which may behave differently when the order of the perturbative expansion
is increased) enter the evaluation of the physical quantities.

In the case of meson DAs~\cite{RQCD:2019osh} the effects related to the
order of the perturbative conversion are of a similar size as those
found here, if one considers the first moments. However, for the second
Gegenbauer moment the effect is more pronounced. Presumably, the second
moments of the baryon DAs are also more sensitive to the
order of the perturbative conversion.

A more recent analysis~\cite{RQCD:2022xux} obtained somewhat different central
values for the low-energy constants that enter the chiral extrapolation: 
$F_0 = 71.1 \, \mathrm{MeV}$, $D=0.57$, $F=0.34$, and
$m_b = 821 \, \mathrm{MeV}$. We repeated our fits using
these new numbers as input (along with the two-loop conversion matrices).
As some results changed by an amount that was considerably
larger than the estimate of the systematic uncertainty due to the chiral
extrapolation determined as in Ref.~\cite{RQCD:2019hps}, we
modified this estimate by adding in quadrature the error evaluated
according to the previous procedure and the difference of the results.
The corresponding numbers can be found in Tables~\ref{tab.results2} and
\ref{tab.moments2}, which are our final results. Table~\ref{tab.results2}
(Table~\ref{tab.moments2}) is to be compared with Table~2 (Table~4) of
Ref.~\cite{RQCD:2019hps}. Notice that due to the scale setting
uncertainty~\cite{RQCD:2022xux} all new results in Tables~\ref{tab.results0}
and \ref{tab.results2} carry an additional error of 1.2\%, while
the scale setting error of the old results in Table~\ref{tab.results0}
amounts to 3\%. The dimensionless quantities displayed in
Tables~\ref{tab.moments0} and \ref{tab.moments2} are not affected by
this uncertainty. 

\begin{acknowledgments}
This work was supported by the Deut\-sche
For\-schungs\-ge\-mein\-schaft (Research Unit FOR 2926 (project~40824754)
and PUNCH4NFDI (project~460248186)). Additional
support from the European Union’s Horizon 2020 research and innovation
programme under the Marie Sk\l{}odowska-Curie grant agreement
no.~813942 (ITN EuroPLEx) and grant agreement no.~824093 (STRONG 2020)
is gratefully acknowledged.

We used a modified version of the {\sc Chroma}~\cite{Edwards:2004sx}
software package along with the {\sc Lib\-Hadron\-Analysis} library
and improved inverters~\cite{Nobile:2010zz,Luscher:2012av,Frommer:2013fsa,
Heybrock:2015kpy}. We thank all our Coordinated Lattice Simulations
(CLS)~\cite{Bruno:2014jqa,Bali:2016umi} colleagues for the joint
generation of the gauge ensembles, using {\sc openQCD}
(\url{https://luscher.web.cern.ch/luscher/openQCD/})~\cite{Luscher:2012av}.

The computation of observables was carried out on the
QPACE~2 and QPACE~3 systems, on the Regensburg
HPC-cluster ATHENE~2, and at various supercomputer centers.  In
particular, the authors gratefully acknowledge computer time granted
by the John von Neumann Institute for Computing (NIC), provided on the
Booster partition of the supercomputer JURECA~\cite{jureca} at
J\"ulich Supercomputing Centre (JSC,
\url{http://www.fz-juelich.de/ias/jsc/}) and granted by the Gauss
Centre for Supercomputing (GCS) on JUWELS~\cite{juwels} at JSC. GCS is
the alliance of the three national supercomputing centres HLRS
(Universit\"at Stuttgart), JSC (For\-schungs\-zen\-trum J\"ulich), and
LRZ (Bayerische Akademie der Wissenschaften), funded by the German
Federal Ministry of Education and Research (BMBF) and the German State
Ministries for Research of Baden-W{\"u}rttemberg (MWK), Bayern
(StMWFK) and Nordrhein-Westfalen (MIWF).
\end{acknowledgments}

\begin{appendix}

\section{Three-quark operator multiplets}
\label{sec.3qops}

In lattice computations it is convenient to employ operator multiplets
that transform irreducibly not only with respect to the lattice symmetry, the
spinorial hypercubic group $\overline{\mathrm {H(4)}}$, but also with
respect to the group $\mathcal S_3$ of permutations of the three quark
flavors. The latter group has three nonequivalent irreducible
representations, which we label by the names of the corresponding
ground state particle multiplets in a flavor symmetric world. Therefore,
the one-dimensional trivial representation is labeled by $\mathscr D$,
the one-dimensional totally antisymmetric representation by $\mathscr S$
and the two-dimensional representation by $\mathscr O$.
Out of the irreducible representations of $\overline{\mathrm {H(4)}}$
only the spinorial representations are relevant for three-quark operators.
They are called $\tau^{\underbar{$\scriptstyle 4$}}_1$,
$\tau^{\underbar{$\scriptstyle 4$}}_2$, $\tau^{\underbar{$\scriptstyle 8$}}$,
$\tau^{\underbar{$\scriptstyle 12$}}_1$,
$\tau^{\underbar{$\scriptstyle 12$}}_2$, where the superscript indicates the
dimension and the subscript distinguishes inequivalent representations
of the same dimension (see, e.g., Ref.~\cite{Kaltenbrunner:2008pb}).

We construct multiplets with the desired transformation properties from
the multiplets defined in Ref.~\cite{Kaltenbrunner:2008pb}. For operators
without derivatives in the representation
$\tau^{\underbar{$\scriptstyle 12$}}_1$ of
$\overline{\mathrm {H(4)}}$ we have one doublet of
operator multiplets transforming according to the two-dimensional
representation of $\mathcal S_3$,
\begin{equation} \label{eq.o12.1}
\mathscr O_1^{\underbar{$\scriptstyle 12$}} = 
\begin{Bmatrix}
\frac{1}{\sqrt{6}}(\cO_7+\cO_8-2\cO_9) \\
\frac{1}{\sqrt{2}}(\cO_7-\cO_8) 
\end{Bmatrix} \,,
\end{equation}
and one operator multiplet transforming trivially under $\mathcal S_3$:
\begin{equation} \label{eq.d12.1}
\mathscr D_1^{\underbar{$\scriptstyle 12$}} =   
\tfrac{1}{\sqrt{3}}(\cO_7+\cO_8+\cO_9) \,.
\end{equation}
For operators without derivatives in the $\overline{\mathrm {H(4)}}$
representation $\tau^{\underbar{$\scriptstyle 4$}}_1$ we have one
multiplet that is totally antisymmetric under flavor permutations,
\begin{equation} \label{eq.s4.1}
\mathscr S_1^{\underbar{$\scriptstyle 4$}} = 
\tfrac{1}{\sqrt{3}}(\cO_3-\cO_4-\cO_5) \,,
\end{equation}
and two doublets of operator multiplets transforming according to the
two-dimensional representation of $\mathcal S_3$:
\begin{align} \label{eq.o4.1a}
( \mathscr O_1^{\underbar{$\scriptstyle 4$}} ) \rule[-1mm]{0mm}{10mm}_1
  =&\begin{Bmatrix}
  \frac{1}{\sqrt{2}}(\cO_3+\cO_4)\\
  \frac{1}{\sqrt{6}}(-\cO_3+\cO_4-2\cO_5)
 \end{Bmatrix}\,,
\\  \label{eq.o4.1b}
( \mathscr O_1^{\underbar{$\scriptstyle 4$}} ) \rule[-1mm]{0mm}{10mm}_2
  =&\begin{Bmatrix}
  \cO_2\\
  \frac{1}{\sqrt{3}}(2\cO_1+\cO_2)
 \end{Bmatrix}\,.
\end{align}

In the case of operators with one derivative we restrict ourselves to 
multiplets that transform according to the $\overline{\mathrm {H(4)}}$
representation $\tau^{\underbar{$\scriptstyle 12$}}_2$, because only these
are safe from mixing with lower-dimensional operators. There are twelve
linearly independent multiplets with this transformation
behavior~\cite{Kaltenbrunner:2008pb}. In Appendix A.2 of
Ref.~\cite{Kaltenbrunner:2008pb} one can find explicit expressions
for four multiplets, labeled $\cO_{D5}$, $\cO_{D6}$, $\cO_{D7}$, $\cO_{D8}$,
where the derivative acts on the third quark field. As the transformation
properties of the operators do not depend on the position of the derivative,
the remaining eight multiplets can be constructed by moving the derivative
to the second or to the first quark field. In the following we replace
the $D$ by $f$, $g$, or $h$ in order to indicate on which quark field the
covariant derivative acts: $f$ ($g$, $h$) means that the derivative acts
on the first (second, third) quark field. 

\begin{widetext}
In this way we get one multiplet that is totally antisymmetric under
$\mathcal S_3$,
\begin{equation} \label{eq.s12.2}
\mathscr S_2^{\underbar{$\scriptstyle 12$}}
  = \tfrac{1}{\sqrt6} \bigl[(\cO_{g5}-\cO_{h5})
  +(\cO_{h6}-\cO_{f6}) +(\cO_{f7}-\cO_{g7})\bigr] \,.
\end{equation}
Additionally there are four doublets of operator multiplets corresponding
to the two-dimensional representation of $\mathcal S_3$,

\begin{align} \label{eq.o12.2a}
(\mathscr O_2^{\underbar{$\scriptstyle 12$}}) \rule[-1mm]{0mm}{10mm}_1
  =&\begin{Bmatrix}
  \frac{1}{3\sqrt2}\bigl[(\cO_{f5}+\cO_{g5}+\cO_{h5})
  +(\cO_{f6}+\cO_{g6}+\cO_{h6})-2(\cO_{f7}
  +\cO_{g7}+\cO_{h7})\bigr]\\
  \frac{1}{\sqrt6}\bigl[(\cO_{f5}+\cO_{g5}+\cO_{h5})
  -(\cO_{f6}+\cO_{g6}+\cO_{h6})\bigr]
 \end{Bmatrix}\,,
\\ \label{eq.o12.2b}
(\mathscr O_2^{\underbar{$\scriptstyle 12$}}) \rule[-1mm]{0mm}{10mm}_2
=&\begin{Bmatrix}
  \frac{1}{6}\bigl[(-2\cO_{f5}+\cO_{g5}+\cO_{h5})
  +(\cO_{f6}-2\cO_{g6}+\cO_{h6})-2(\cO_{f7}
  +\cO_{g7}-2\cO_{h7})\bigr]\\
  \frac{1}{2\sqrt3}\bigl[(-2\cO_{f5}+\cO_{g5}
  +\cO_{h5})-(\cO_{f6}-2\cO_{g6}+\cO_{h6})\bigr]
 \end{Bmatrix}\,,
\\ \label{eq.o12.2c}
(\mathscr O_2^{\underbar{$\scriptstyle 12$}}) \rule[-1mm]{0mm}{10mm}_3
=&\begin{Bmatrix}
  \frac{1}{2}\bigl[(\cO_{g5}-\cO_{h5})-(\cO_{h6}
  -\cO_{f6})\bigr]\\
  \frac{1}{2\sqrt3}\bigl[(\cO_{h5}-\cO_{g5})
  +(\cO_{f6}-\cO_{h6})-2(\cO_{g7}
  -\cO_{f7})\bigr]
 \end{Bmatrix}\,,
\\ \label{eq.o12.2d}
(\mathscr O_2^{\underbar{$\scriptstyle 12$}}) \rule[-1mm]{0mm}{10mm}_4
=&\begin{Bmatrix}
  \frac{1}{\sqrt6}(\cO_{f8}+\cO_{g8}-2\cO_{h8})\\
  \frac{1}{\sqrt2}(\cO_{f8}-\cO_{g8})
 \end{Bmatrix}\,,
\end{align}
and three operator multiplets transforming trivially under flavor permutations:
\begin{align} \label{eq.d12.2a}
(\mathscr D_2^{\underbar{$\scriptstyle 12$}}) \rule[-1mm]{0mm}{5mm}_1
= {} &\tfrac{1}{3}\bigl[(\cO_{f5}+\cO_{g5}+\cO_{h5})
  +(\cO_{f6}+\cO_{g6}+\cO_{h6})+(\cO_{f7}
  +\cO_{g7}+\cO_{h7})\bigr] \,, \\  \label{eq.d12.2b}
(\mathscr D_2^{\underbar{$\scriptstyle 12$}}) \rule[-1mm]{0mm}{5mm}_2
= {} &\tfrac{1}{3\sqrt2}\bigl[(-2\cO_{f5}+\cO_{g5}
  +\cO_{h5})+(\cO_{f6}-2\cO_{g6}+\cO_{h6})
  +(\cO_{f7}+\cO_{g7}-2\cO_{h7})\bigr] \,,
  \\ \label{eq.d12.2c}
(\mathscr D_2^{\underbar{$\scriptstyle 12$}}) \rule[-1mm]{0mm}{5mm}_3
= {} &\tfrac{1}{\sqrt3}(\cO_{f8}+\cO_{g8}+\cO_{h8}) \,.
\end{align}
\end{widetext}

\section{Conversion matrices for lattice operators}
\label{sec.converlat}

In this Appendix we collect numerical values for the two-loop conversion
matrices of the three-quark operators used in the lattice calculations.
They refer to the momentum configuration employed in our renormalization
condition on the lattice, where we used the following momenta (in
Euclidean space) for the three external quark lines:
\begin{equation}
\begin{split}
p_1 & {} = \frac{\mu}{2} (+1,+1,+1,+1) \,, \\
p_2 & {} = \frac{\mu}{2} (-1,-1,-1,+1) \,, \\
p_3 & {} = \frac{\mu}{2} (+1,-1,-1,-1) \,.
\end{split}
\end{equation}
Notice that the scale $\mu$ will drop out in the coefficients of the
perturbative expansion of the conversion matrices.
Analytical expressions in the one-loop approximation can be found in
Appendix~G of Ref.~\cite{RQCD:2020kuu}.

With the $\MS$ coupling constant $\bar{g}$ we write the two-loop
conversion matrix $C_{mm'}$ as
\begin{equation}
C_{mm'} = \delta_{mm'} + \frac{\bar{g}^2}{16 \pi^2} (c_1)_{mm'}
  + \left( \frac{\bar{g}^2}{16 \pi^2} \right)^2 (c_2)_{mm'} \,.
\end{equation}

For the multiplets
$\mathscr O_1^{\underbar{$\scriptstyle 12$}}$ and
$\mathscr D_1^{\underbar{$\scriptstyle 12$}}$ 
we have
\begin{equation}
C_{11}(\mathscr O_1^{\underbar{$\scriptstyle 12$}}) =
C_{11}(\mathscr D_1^{\underbar{$\scriptstyle 12$}})
\end{equation}
with 
\begin{equation}
\begin{split}
(c_1)_{11} &= - 0.0493633154  \,, \\
(c_2)_{11} &= - 38.45080(90) + 3.746206(21) \, n_f \,.
\end{split}
\end{equation}
In the case of the multiplet $\mathscr S_1^{\underbar{$\scriptstyle 4$}}$
one finds
\begin{equation}
\begin{split}
(c_1)_{11} &= 2.629711269  \,, \\
(c_2)_{11} &= 4.3294(24) + 0.004027(37) \, n_f \,.
\end{split}
\end{equation}
The multiplets $\mathscr O_1^{\underbar{$\scriptstyle 4$}}$ have a 
diagonal $2 \times 2$ mixing matrix with
\begin{equation}
\begin{split}
(c_1)_{11} &= 2.629711269  \,, \\
(c_2)_{11} &= 4.3294(24) + 0.004027(37) \, n_f \,, \\
(c_1)_{22} &= 2.629711269  \,, \\
(c_2)_{22} &= 1.3116(24) + 0.004027(37) \, n_f \,.
\end{split}
\end{equation}
For the multiplet $\mathscr S_2^{\underbar{$\scriptstyle 12$}}$ of operators
with one derivative the conversion factor is given by
\begin{equation}
\begin{split}
(c_1)_{11} &= -3.376398061  \,, \\
(c_2)_{11} &= -105.555(12) + 11.076863(17) \, n_f \,.
\end{split}
\end{equation}
The multiplets $\mathscr O_2^{\underbar{$\scriptstyle 12$}}$
have a $4 \times 4$ mixing matrix with the diagonal entries
\begin{equation}
\begin{split}
(c_1)_{11} &= 0.05197412907 \,, \\
(c_2)_{11} &= -34.599(22) + 35.220102(34) \, n_f \,, \\
(c_1)_{22} &= -3.777434416 \,, \\
(c_2)_{22} &= -109.878(12) + 11.563987(18) \, n_f \,, \\
(c_1)_{33} &= -3.376398061 \,, \\
(c_2)_{33} &= -105.555(12) + 11.076863(17) \, n_f \,, \\
(c_1)_{44} &= -4.450577872 \,, \\
(c_2)_{44} &= -114.543(13) + 12.620297(22) \, n_f \,.
\end{split}
\end{equation}
The only nonvanishing off-diagonal entries are
\begin{equation}
\begin{split}
(c_1)_{12} &= 0.05875235294 \,, \\
(c_2)_{12} &= 1.8382(48) - 0.1089969(66) \, n_f \,, \\
(c_1)_{21} &= 0.2350094118 \,, \\
(c_2)_{21} &= 7.858(10) - 0.4359875(86) \, n_f \,.
\end{split}
\end{equation}
In the case of the multiplets $\mathscr D_2^{\underbar{$\scriptstyle 12$}}$
we get a $3 \times 3$ mixing matrix, whose nonzero entries are given by
\begin{equation}
C_{mm'}(\mathscr D_2^{\underbar{$\scriptstyle 12$}}) =
C_{mm'}(\mathscr O_2^{\underbar{$\scriptstyle 12$}})
\end{equation}
for $m,m' \in \{1,2\}$ and
\begin{equation}
\begin{split}
(c_1)_{33} &= -1.388900608  \,, \\
(c_2)_{33} &= -56.736(25) + 5.617296(39) \, n_f \,.
\end{split}
\end{equation}

The uncertainties of the two-loop coefficients arising from the numerical
evaluation of the master intergrals have been estimated by error propagation.
Since in this procedure possible correlations have been neglected, the given
errors are most likely overestimated.
  
\section{Anomalous dimensions for general operators}
\label{sec.anodimgen}
Every local three-quark operator can be represented as a linear
combination of the operators
\begin{equation} 
\epsilon_{i_1 i_2 i_3} (D_{\bar{l}_1} \psi^{f_1} (x))^{i_1}_{\alpha_1}
(D_{\bar{l}_2} \psi^{f_2} (x))^{i_2}_{\alpha_2}
(D_{\bar{l}_3} \psi^{f_3} (x))^{i_3}_{\alpha_3} \,.
\end{equation} 
Here we use a multi-index notation for the covariant derivatives,
$D_{\bar{l}} \equiv D_{\lambda_1} \cdots  D_{\lambda_l}$.
The indices $\alpha_1$, {\ldots} are spinor indices, the indices
$i_1$, {\ldots} are color indices, and the flavor indices $f_1$, {\ldots}
take the values 1,2,3 where $\psi^1 = u$, $\psi^2 = d$, $\psi^3 = s$. 
In the context of renormalization all flavors are taken to be massless.
In the following all quantities are to be understood as Euclidean.

The basic object for the renormalization of the local three-quark operators
is the ``flavorless'' amputated four-point function
\begin{widetext}
\begin{equation} \label{eq.fourpt}
\begin{split}
H^{\beta_1 \beta_2 \beta_3}_{\alpha_1 \alpha_2 \alpha_3}
&(\bar{l}_1, \bar{l}_2, \bar{l}_3;p_1, p_2, p_3) =
- \int \!  d^4x_1 \, d^4x_2 \, d^4x_3 \, 
e^{i (p_1 \cdot x_1 + p_2 \cdot x_2 + p_3 \cdot x_3)}
\epsilon_{j_1 j_2 j_3} \epsilon_{i_1 i_2 i_3}
\\ & {} \times 
\langle ( D_{\bar{l}_1} u (0))^{j_1}_{\beta_1} 
( D_{\bar{l}_2} d (0))^{j_2}_{\beta_2} 
( D_{\bar{l}_3} s (0))^{j_3}_{\beta_3} 
\bar{u}^{i_1}_{\alpha'_1} (x_1) \bar{d}^{i_2}_{\alpha'_2} (x_2)
\bar{s}^{i_3}_{\alpha'_3} (x_3) \rangle
G_2^{-1}(p_1)_{\alpha'_1 \alpha_1} G_2^{-1}(p_2)_{\alpha'_2 \alpha_2}
G_2^{-1}(p_3)_{\alpha'_3 \alpha_3} \,.
\end{split}
\end{equation}
\end{widetext}
Here $\langle \cdots \rangle$ indicates the functional integral with
the gauge fields fixed, e.g., to Landau gauge; $p_1$, $p_2$, $p_3$ are the
external momenta. The two-point function $G_2 (p)$ required for the
amputation is defined by
\begin{equation} 
G_2 (p)_{\alpha' \alpha} \delta_{ij} = \int \! d^4x \, e^{i p \cdot x}
\langle u^j_{\alpha'} (0) \bar{u}^i_\alpha (x) \rangle \,.
\end{equation}

In the case of operators without derivatives, the renormalized (in the
$\MS$-like scheme suggested in Ref.~\cite{Krankl:2011gch})
four-point function $\bar{H}^{\beta_1 \beta_2 \beta_3}_
{\alpha_1 \alpha_2 \alpha_3} (-,-,-;p_1,p_2,p_3)$
is related to its bare counterpart by
\begin{equation} \label{eq.reno0d}
\begin{split}
\bar{H}&^{\beta_1 \beta_2 \beta_3}_{\alpha_1 \alpha_2 \alpha_3}
(-,-,-;p_1,p_2,p_3) \\ & {} =
Z_q^{-3/2} Z^{\beta_1 \beta_2 \beta_3}_{\beta'_1 \beta'_2 \beta'_3}
H^{\beta'_1 \beta'_2 \beta'_3}_{\alpha_1 \alpha_2 \alpha_3}
 (-,-,-;p_1,p_2,p_3) \,.
\end{split}
\end{equation}
The matrix of anomalous dimensions can then be calculated as 
\begin{equation} 
\gamma = - \left( \mu \frac{d Z}{d \mu} \right) Z^{-1} \,.
\end{equation}
Here $\mu$ is the renormalization scale and $Z_q$ is the wave function
renormalization constant of the quark fields. With the $\MS$ coupling
constant $\bar{g}$ the anomalous dimension can be written as
\begin{equation}
\gamma = \gamma_0 \frac{\bar{g}^2}{16 \pi^2} 
  + \gamma_1 \left( \frac{\bar{g}^2}{16 \pi^2} \right)^2 + \cdots \,.
\end{equation}
The one-loop coefficient is given by
\begin{equation} \label{eq.anodim0}
\gamma_0 =  - \frac{1}{3} ( G_{022} + G_{202} + G_{220}) \,.
\end{equation}
The objects $G_{022}$ etc.\ carry six spinor indices like $Z$ and are
defined as 
\begin{equation} 
\begin{split}
G_{022} &= \Gamma_0 \otimes \Gamma_{\mu_1 \mu_2} \otimes \Gamma_{\mu_1 \mu_2}
\,, \\
G_{202} &= \Gamma_{\mu_1 \mu_2} \otimes \Gamma_0 \otimes \Gamma_{\mu_1 \mu_2}
\,, \\
G_{220} &= \Gamma_{\mu_1 \mu_2} \otimes \Gamma_{\mu_1 \mu_2} \otimes \Gamma_0 
\end{split}
\end{equation}
in terms of the totally antisymmetric products of $\gamma$ matrices
\begin{equation} 
\Gamma_0 = \mathds{1} \,, \; \Gamma_{\mu \nu} = \frac{1}{2!}
\left( \gamma_\mu \gamma_\nu - \gamma_\nu \gamma_\mu \right) \,, \ldots
\end{equation}
Three-loop expressions for the anomalous dimensions of the operators without
derivatives are given in Ref.~\cite{Gracey:2012gx}.

For the operators with one derivative the renormalization structure is
more complicated. In addition to the matrix structure (encoded in the
$G_{lmn}$) that operates on the open spinor indices, we have a $3 \times 3$
matrix structure, because the operators with the covariant derivative acting
on the first, second, or third quark field mix with each other. Instead of
Eq.~\eqref{eq.reno0d} we now have
\begin{widetext}
\begin{equation} 
\begin{pmatrix}
\bar{H}^{\beta_1 \beta_2 \beta_3}_{\alpha_1 \alpha_2 \alpha_3}
(\mu,-,-;p_1,p_2,p_3) \\
\bar{H}^{\beta_1 \beta_2 \beta_3}_{\alpha_1 \alpha_2 \alpha_3}
(-,\mu,-;p_1,p_2,p_3) \\
\bar{H}^{\beta_1 \beta_2 \beta_3}_{\alpha_1 \alpha_2 \alpha_3}
(-,-,\mu;p_1,p_2,p_3)  \end{pmatrix} = Z_q^{-3/2} 
Z^{\beta_1 \beta_2 \beta_3}_{\beta'_1 \beta'_2 \beta'_3}
\begin{pmatrix}
H^{\beta'_1 \beta'_2 \beta'_3}_{\alpha_1 \alpha_2 \alpha_3}
(\mu,-,-;p_1,p_2,p_3) \\
H^{\beta'_1 \beta'_2 \beta'_3}_{\alpha_1 \alpha_2 \alpha_3}
(-,\mu,-;p_1,p_2,p_3) \\
H^{\beta'_1 \beta'_2 \beta'_3}_{\alpha_1 \alpha_2 \alpha_3}
(-,-,\mu;p_1,p_2,p_3)  \end{pmatrix} \,,
\end{equation}
where $Z^{\beta_1 \beta_2 \beta_3}_{\beta'_1 \beta'_2 \beta'_3}$ is a
$3 \times 3$ matrix. The one-loop contribution to the matrix of anomalous
dimensions for operators of leading twist is given by
\begin{equation} \label{eq.anodim1}
\begin{split}
\gamma_0 &= \begin{pmatrix} 32/9 & -16/9 & -16/9 \\
  -16/9 & 32/9 & -16/9 \\ -16/9 & -16/9 & 32/9 \end{pmatrix} \otimes \, G_{000}
+ \begin{pmatrix} -1/3 & 0 & 0 \\ 0 & -2/9 & -1/9 \\
  0 & -1/9 & -2/9 \end{pmatrix} \otimes \, G_{022}
+ \begin{pmatrix} -2/9 & 0 & -1/9 \\ 0 & -1/3 & 0 \\
  -1/9 & 0 & -2/9 \end{pmatrix} \otimes \, G_{202}
\\ & {}
+ \begin{pmatrix} -2/9 & -1/9 & 0 \\ -1/9 & -2/9 & 0 \\
  0 & 0 & -1/3 \end{pmatrix} \otimes \, G_{220} \,.
\end{split}
\end{equation}
The two-loop contribution, again projected onto leading twist, reads
\begin{equation}  \label{eq.anodim2}
\begin{split}
\gamma_1 &= \begin{pmatrix}
8348/81 - (508/81)n_f & & - 1339/81 + (92/81)n_f & & - 1339/81 + (92/81)n_f \\
- 1339/81 + (92/81)n_f & & 8348/81 - (508/81)n_f & & - 1339/81 + (92/81)n_f \\
- 1339/81 + (92/81)n_f & & - 1339/81 + (92/81)n_f & & 8348/81 - (508/81)n_f 
\end{pmatrix} \otimes \, G_{000}
\\ & {} + \begin{pmatrix}
- 1265/486 + (1/27)n_f & & - 55/243 & & -55/243 \\
- 29/486 & & - 817/486 + (4/81)n_f & & - 22/27 - (1/81)n_f \\
- 29/486 & & - 22/27 - (1/81)n_f & & - 817/486 + (4/81)n_f
\end{pmatrix} \otimes \, G_{022}
\\ & {} + \begin{pmatrix}
- 817/486 + (4/81)n_f & & - 29/486 & & - 22/27 - (1/81)n_f \\
- 55/243 & & - 1265/486 + (1/27)n_f & & - 55/243 \\
- 22/27 - (1/81)n_f & & - 29/486 & & - 817/486 + (4/81)n_f 
\end{pmatrix} \otimes \, G_{202}
\\ & {} + \begin{pmatrix}
- 817/486 + (4/81)n_f & & - 22/27 - (1/81)n_f & & - 29/486 \\
- 22/27 - (1/81)n_f & & - 817/486 + (4/81)n_f & & - 29/486 \\
- 55/243 & & - 55/243 & & - 1265/486 + (1/27)n_f 
\end{pmatrix} \otimes \, G_{220}
\\ & {} + \begin{pmatrix}
1/9 & & 0 & &  0 \\
0 & & 109/1944 & & 107/1944 \\
0 & & 107/1944 & & 109/1944
\end{pmatrix} \otimes \, G_{044}
+ \begin{pmatrix}
109/1944 & & 0 & & 107/1944 \\
0 & & 1/9 & & 0 \\ 
107/1944 & & 0 & & 109/1944 
\end{pmatrix} \otimes \, G_{404}
\\ & {} + \begin{pmatrix}
109/1944 & & 107/1944 & & 0 \\
107/1944 & & 109/1944 & & 0 \\
0 & & 0 & & 1/9
\end{pmatrix} \otimes \, G_{440}
\\ & {} + \begin{pmatrix}
- 4/243 & & - 5/486 & & - 5/486 \\
- 19/972 & & - 1/27 & & - 2/243 \\
- 19/972 & & - 2/243& & - 1/27 
\end{pmatrix} \otimes \, G_{422}
+ \begin{pmatrix}
- 1/27 & & - 19/972 & & - 2/243 \\
- 5/486 & & - 4/243 & & - 5/486 \\
- 2/243 & & - 19/972 & & - 1/27
\end{pmatrix} \otimes \, G_{242}
\\ & {} + \begin{pmatrix}
- 1/27 & & - 2/243 & & - 19/972 \\
- 2/243 & &  - 1/27 & & - 19/972 \\
- 5/486 & & - 5/486 & & - 4/243
\end{pmatrix} \otimes \, G_{224}
+ \begin{pmatrix}
0 & & 2/81 & & - 2/81 \\
- 2/81 & &  0 & & 2/81 \\
2/81 & & - 2/81 & & 0
\end{pmatrix} \otimes \, G_{222} \,.
\end{split}
\end{equation}
\end{widetext}

The one-loop anomalous dimension matrix \eqref{eq.anodim1} was first
presented in this form in Ref.~\cite{Gruber:2017ozo}, unfortunately with
some typos. However, the analysis code employed in
Refs.~\cite{RQCD:2019hps,Bali:2015ykx,RQCD:2020kuu} was correct.
The two-loop result in \eqref{eq.anodim2} is new. Notice that the anomalous
dimensions \eqref{eq.anodim0}, \eqref{eq.anodim1} and \eqref{eq.anodim2}
are not only relevant for spin~$1/2$ baryons, but also for spin~$3/2$ baryons.

\section{Anomalous dimensions for lattice operators}
\label{sec.anodimlat}

The lattice operators (or rather operator multiplets) that are relevant for the
evaluation of (moments of) baryon DAs can be found in Appendix~\ref{sec.3qops}.
In Appendix~E of Ref.~\cite{RQCD:2020kuu} the three-loop
anomalous dimensions of the operators without derivatives are given,
derived from the results in Ref.~\cite{Gracey:2012gx}.
Here we present the new two-loop anomalous dimensions of the operators
with one derivative.

For the multiplet $\mathscr S_2^{\underbar{$\scriptstyle 12$}}$ we get
\begin{equation}
\begin{split}
\gamma_0 = {} & \frac{52}{9} \,, \\
\gamma_1 = {} & \frac{28990}{243} - \frac{620}{81} n_f \,.
\end{split}
\end{equation}

In the case of the four mixing multiplets 
$(\mathscr O_2^{\underbar{$\scriptstyle 12$}}) \rule[-1mm]{0mm}{5mm}_1, \ldots
,(\mathscr O_2^{\underbar{$\scriptstyle 12$}}) \rule[-1mm]{0mm}{5mm}_4$
one finds
\begin{widetext}
\begin{equation}
\begin{split}
\gamma_0 &= \begin{pmatrix}
4/3 & & 0 & & 0 & & 0 \\   
0  & & 20/3 & & 0 & & 0 \\   
0  & & 0 & & 52/9 & & 0 \\   
0 & & 0 & & 0 & & 8 \end{pmatrix} \,, \\
\gamma_1 &= \begin{pmatrix}
236/3 - 112 n_f/27  & & - 16 \sqrt{2}/27  & & 0 & & 0 \\
0 & & 1174/9 - 68 n_f/9 & & 0 & & 0 \\
0 & & 0 & & 28990/243 - 620 n_f/81  & & 0 \\
0 & & 0 & & 0 & & 428/3 - 8 n_f \end{pmatrix} \,.
\end{split}
\end{equation}

Finally, we have the three mixing multiplets
$(\mathscr D_2^{\underbar{$\scriptstyle 12$}}) \rule[-1mm]{0mm}{5mm}_1, \ldots
,(\mathscr D_2^{\underbar{$\scriptstyle 12$}}) \rule[-1mm]{0mm}{5mm}_3$
with
\begin{equation}
\begin{split}
\gamma_0 &= \begin{pmatrix}
4/3 & & 0 & & 0 \\   
0  & & 20/3 & & 0 \\   
0 & & 0 & & 4 \end{pmatrix} \,, \\
\gamma_1 &= \begin{pmatrix}
236/3 - 112 n_f/27 & & - 16 \sqrt{2}/27 & & 0 \\
0 & & 1174/9 - 68 n_f/9 & & 0 \\
0 & & 0 & & 328/3 - 40 n_f/9 \end{pmatrix} \,.
\end{split}
\end{equation}
\end{widetext}

\section{Comparing one-loop and two-loop conversion}
\label{sec.compare}

\begin{figure}[h]
\includegraphics[width=.45\textwidth]{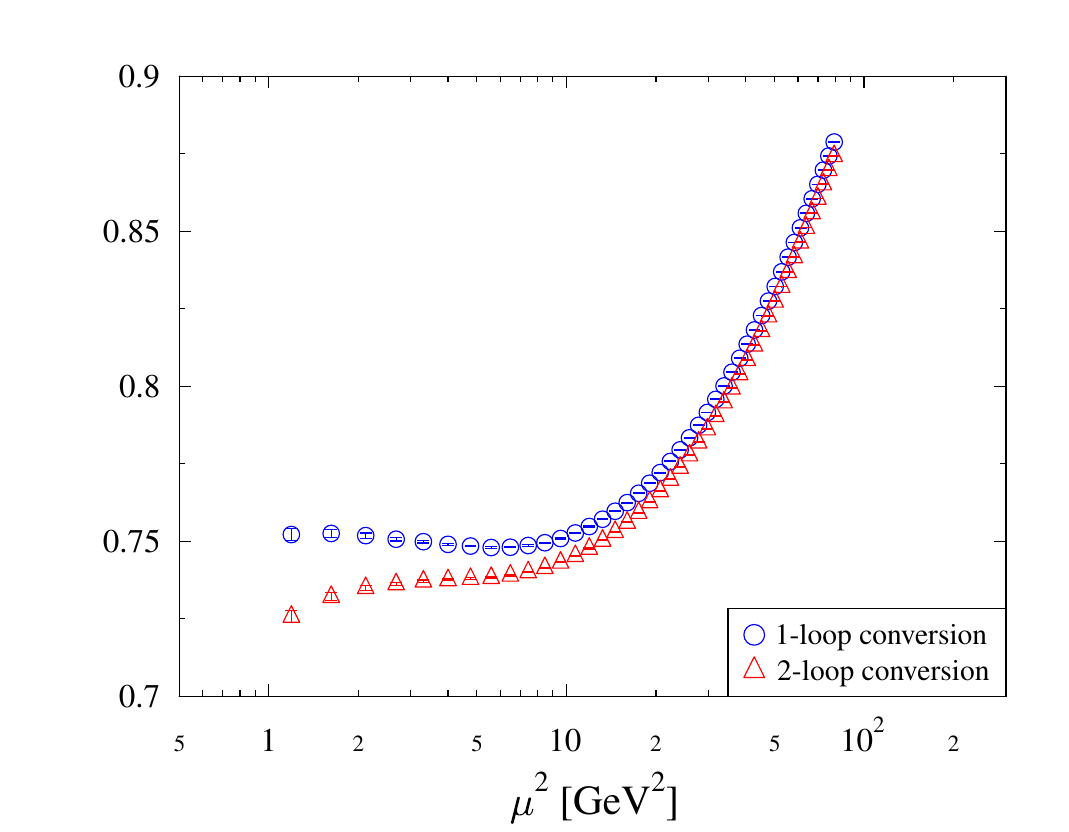}
\includegraphics[width=.45\textwidth]{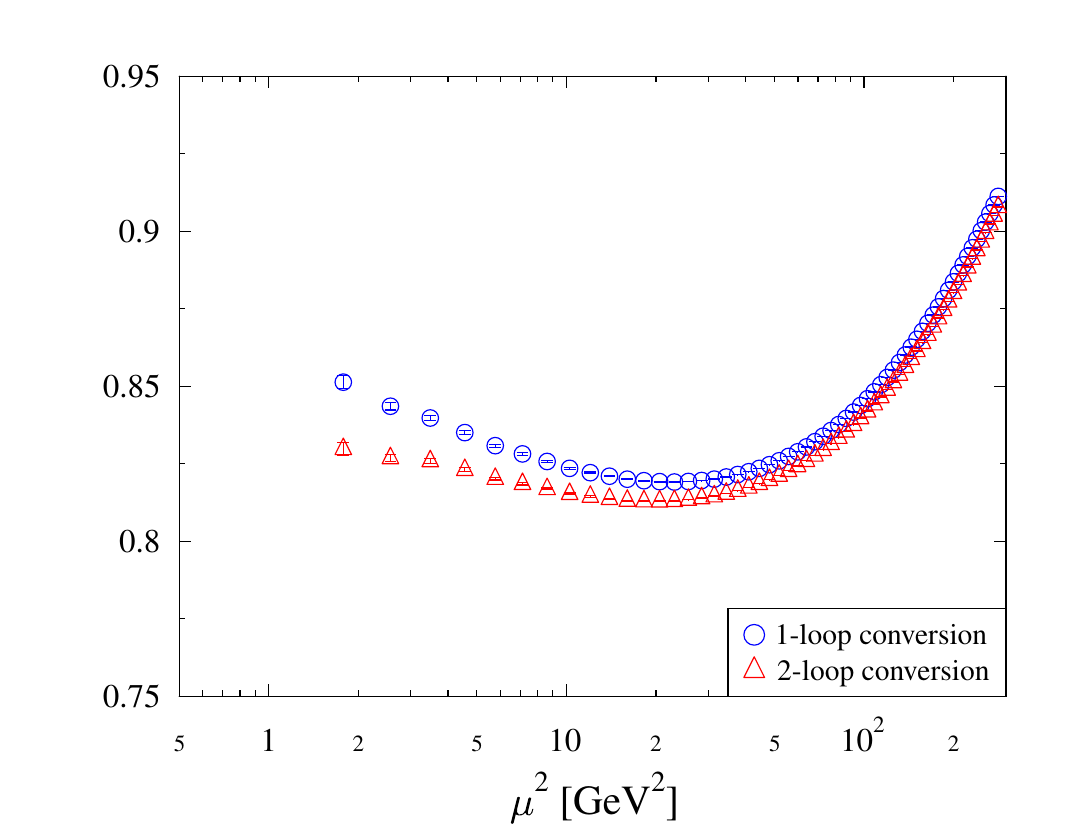}
\caption{\label{fig.comp0d1} Renormalization factor for the multiplet
$\mathscr O_1^{\underbar{$\scriptstyle 12$}}$ rescaled to the target
scale of 2 GeV for $\beta=3.4$ (top panel) and $\beta=3.85$ (bottom panel).}
\end{figure}

\begin{figure}[h]
\includegraphics[width=.45\textwidth]{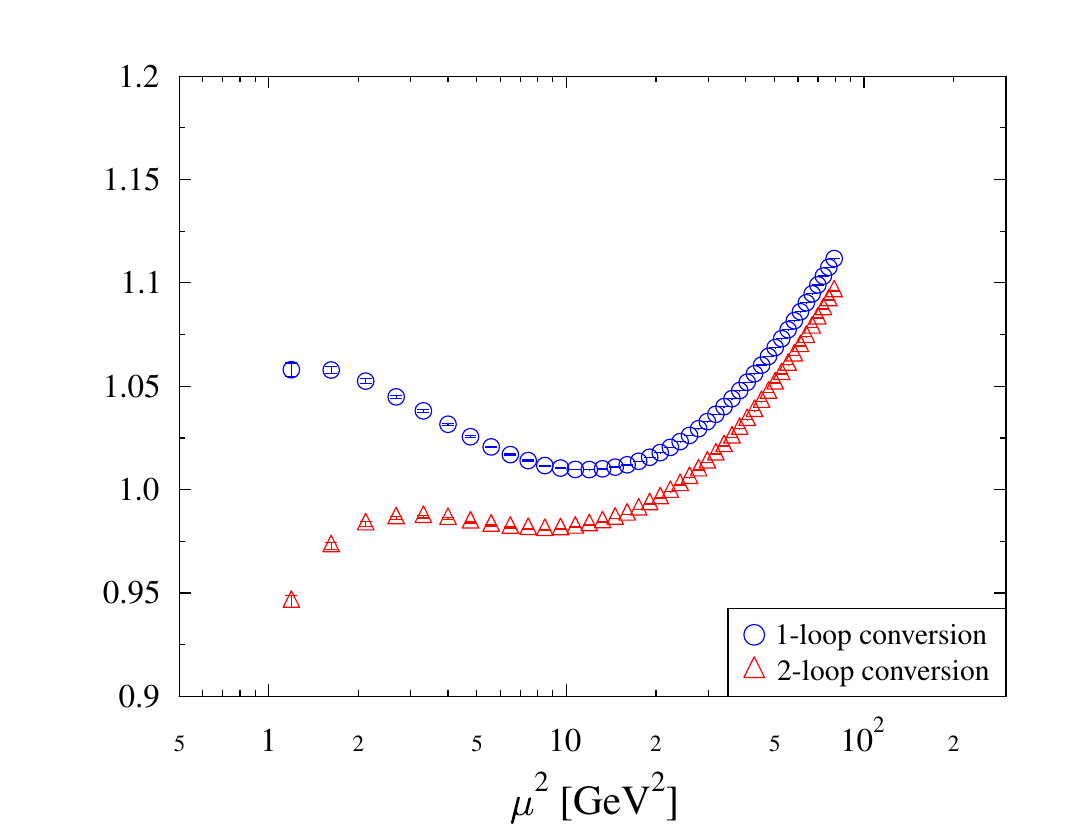}
\includegraphics[width=.45\textwidth]{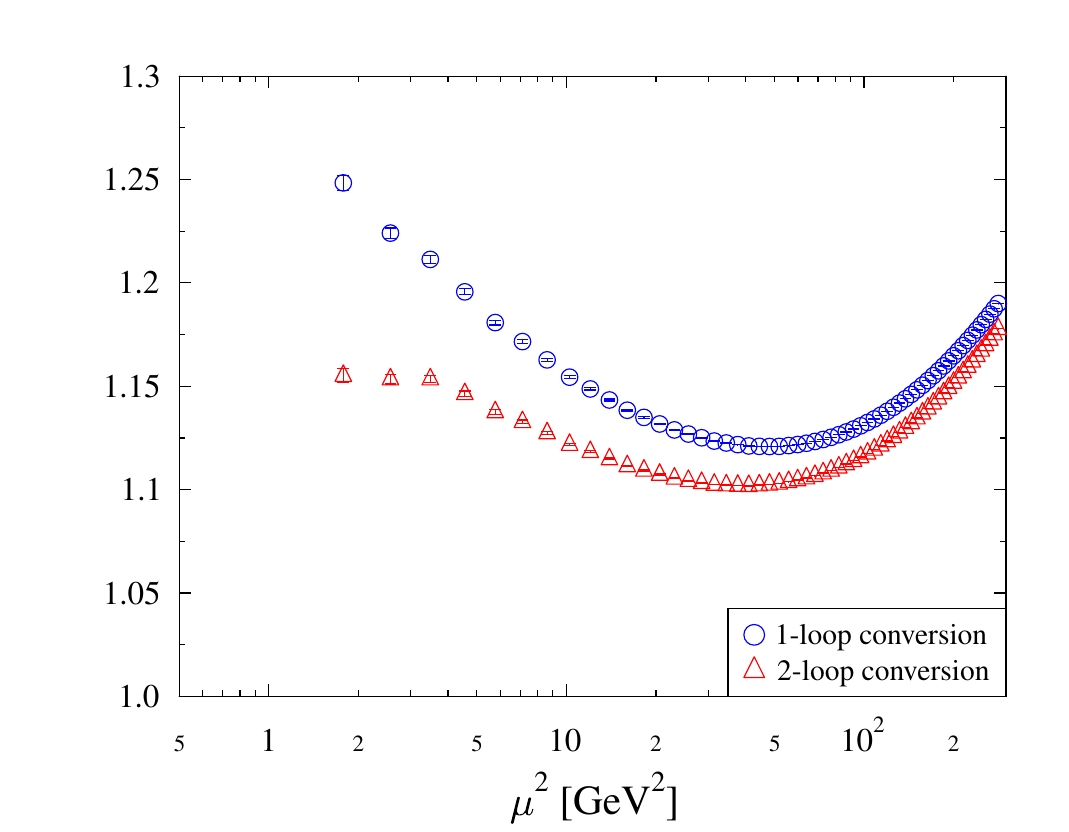}
\caption{\label{fig.comp1d16} Renormalization factor for the multiplet
$\mathscr S_2^{\underbar{$\scriptstyle 12$}}$ rescaled to the target
scale of 2 GeV for $\beta=3.4$ (top panel) and $\beta=3.85$ (bottom panel).}
\end{figure}

In this Appendix we present some plots directly comparing renormalization
factors obtained with one-loop and two-loop conversion. We do this for the
multiplets $\mathscr O_1^{\underbar{$\scriptstyle 12$}}$
(Fig.~\ref{fig.comp0d1}) and $\mathscr S_2^{\underbar{$\scriptstyle 12$}}$
(Fig.~\ref{fig.comp1d16}), separately for our coarsest 
($a^{-1} \approx 2.3 \, \mathrm{GeV}$, $\beta=3.4$) and finest 
($a^{-1} \approx 5.1 \, \mathrm{GeV}$, $\beta=3.85$) lattices.
Notice that these data are included in Figs.~\ref{fig.zfac0d1} and
\ref{fig.zfac1d16}.

Replacing one-loop by two-loop conversion leads to larger effects
for the multiplet $\mathscr S_2^{\underbar{$\scriptstyle 12$}}$ of operators
with one derivative than for the multiplet
$\mathscr O_1^{\underbar{$\scriptstyle 12$}}$ of operators without derivatives. 
Still, in the relevant range of scales around $\mu^2 = 4 \, \mathrm{GeV}^2$
these effects are of moderate size ($\sim 1.5 \%$ for
$\mathscr O_1^{\underbar{$\scriptstyle 12$}}$ and $\sim 4.5 \%$ for
$\mathscr S_2^{\underbar{$\scriptstyle 12$}}$), and the uncertainties due to
the perturbative conversion are not the major source of systematic errors
of our final results.

\end{appendix}

\end{document}